%% file: 0.main.tex
\documentclass[preprint, 5p, times, 10pt, lefttitle]{elsarticle}
\usepackage{hyperref}
\usepackage[switch, modulo]{lineno}
\biboptions{sort&compress}
\usepackage{mdframed} 
\usepackage{multicol}
\usepackage{nomencl}
\renewcommand*\nompreamble{\begin{multicols}{2}}
\renewcommand*\nompostamble{\end{multicols}}
\makenomenclature
\usepackage{etoolbox}
\renewcommand\nomgroup[1]{%
  \item[\bfseries
  \ifstrequal{#1}{I}{Indices}{%
  \ifstrequal{#1}{P}{Parameters}{%
  \ifstrequal{#1}{V}{Variables}{%
  \ifstrequal{#1}{A}{Acronyms}
  {}}}}%
]}
\usepackage{array}
\usepackage{makecell}
\usepackage{tabularx}
\usepackage{float}
\usepackage{textcomp}
\usepackage{amsmath}
\usepackage{subcaption}
\usepackage{amssymb}
\usepackage{threeparttable}

\begin{document}

\begin{frontmatter}

\title{Distributionally Robust Chance-Constrained Flexibility Planning for Integrated Energy System}

\author[tue]{Sen Zhan}
\ead{s.zhan@tue.nl}
\author[sewpg]{Peng Hou\corref{cor}}
\cortext[cor]{Corresponding author}
\ead{houpeng@shanghai-electric.com}
\author[dtu]{Guangya Yang}
\ead{gyy@elektro.dtu.dk}
\address[tue]{Electrical Energy Systems, Department of Electrical Engineering, Eindhoven University of Technology, 5612 AZ Eindhoven, the Netherlands}
\address[sewpg]{SEWPG European Innovation Center, 8000 Aarhus, Denmark}
\address[dtu]{Department of Electrical Engineering, Technical University of Denmark, 2800 Kgs. Lyngby, Denmark}

\begin{abstract}
Inflexible combined heat and power (CHP) plants and uncertain wind power production result in excess power in distribution networks, which leads to inverse power flow challenging grid operations. Power-to-X facilities such as electrolysers and electric boilers can offer extra flexibility to the integrated energy system. In this regard, we aim to jointly determine the optimal Power-to-X facility sizing and integrated energy system operations in this study. To account for wind power uncertainties, a distributionally robust chance-constrained model is developed to characterize wind power uncertainties using ambiguity sets. Linear decision rules are applied to analytically express real-time recourse actions when uncertainties are exposed, which allows the propagation of wind power uncertainties to gas and heat systems. Accordingly, the developed three-stage distributionally robust chance-constrained model is converted into a computationally tractable single-stage mixed-integer conic model. A case study validates the effectiveness of introducing the electrolyser and electric boiler into the integrated energy system, with respect to the decreased system cost, expanded CHP plant flexibility and reduced inverse power flow. The developed distributionally robust optimization model exhibits better effectiveness and robustness compared to a chance-constrained optimization model assuming wind forecast errors follow Gaussian distribution. Detailed profit analysis reveals that although the overall system cost is minimized, the profit is distributed unevenly across various stakeholders in the system. The profit mainly falls with the wind power plants, which therefore are most motivated to make investments in the flexibility resources. Other parties rely on additional policies such as bilateral contracts with the wind power plants to gain incentives to invest. The findings from this study can be used to motivate policy-makers to make proper regulations to incentivize investments in flexibility resources and establish a more reliable power grid.
\end{abstract}

\end{frontmatter}

\begin{table*}[htbp]
\vspace{-1.5cm}
\begin{mdframed}
\input{nomenclature}
\printnomenclature
\end{mdframed}
\end{table*}

\input{1.intro}

\input{2.components}
\input{3.MathModel}

\input{4.Case}

\input{5.Results}

\input{6.Conclusions}
\input{7.appendix}

\bibliographystyle{elsarticle-num}
\bibliography{reference}

\end{document}

%% file: nomenclature.tex
\nomenclature[I]{$t$}{Hour index}
\nomenclature[I]{$r$}{Representative day index}
\nomenclature[I]{$k$}{CHP corner point index}
\nomenclature[I]{$i$}{P2HH corner point index}
\nomenclature[I]{$j$}{Auxiliary binary variable index}

\nomenclature[A]{IES}{Integrated Energy System}
\nomenclature[A]{AEC}{Alkaline Electrolyser}
\nomenclature[A]{P2H}{Power to Hydrogen}
\nomenclature[A]{P2HH}{Power to Hydrogen and Heat}
\nomenclature[A]{CHP}{Combined Heat and Power}
\nomenclature[A]{SOC}{Second Order Cone}
\nomenclature[A]{DRCC}{Distributionally Robust Chance-Constrianed}
\nomenclature[A]{CC}{Chance-Constrained}
\nomenclature[A]{EB}{Electric Boiler}
\nomenclature[A]{RO}{Robust Optimization}
\nomenclature[A]{ARO}{Adaptive Robust Optimization}
\nomenclature[A]{MIQCP}{Mixed-Integer Quadratically Constrained Programming}
\nomenclature[A]{MILP}{Mixed-Integer Linear Programming}
\nomenclature[A]{MISOCCQP}{Mixed-Integer Second-Order Cone  Constrained Quadratic Programming}
\nomenclature[A]{HCNG}{Hydrogen-enriched Compressed Natural Gas}
\nomenclature[A]{AC}{Absorption Chiller}
\nomenclature[A]{GT}{Gas Turbine}
\nomenclature[A]{TT}{Thermal storage Tank}
\nomenclature[A]{BT}{Battery}
\nomenclature[A]{SHS}{Seasonal Hydrogen Storage}
\nomenclature[A]{PEMEC}{Polymer Electrolyte Membrane Electrolyser}
\nomenclature[A]{SOEC}{Solid Oxide Electrolyser}

\nomenclature[P]{$F$}{Faraday's constant}
\nomenclature[P]{$c^{el}$}{Annualized single electrolyser cell cost}
\nomenclature[P]{$c^{su}, c^{sd}$}{Start-up/shut-down cost of CHP plant}
\nomenclature[P]{$\mathbf{\mu}$}{Mean of prediction error of distributed wind generators}
\nomenclature[P]{$\mathbf{\Sigma}$}{Covariance of prediction error of distributed wind generators}
\nomenclature[P]{$RU, RD$}{Ramp up/down rate of CHP plant}
\nomenclature[P]{$SU, SD$}{Start-up/shut-down rate of CHP plant}
\nomenclature[P]{$k_r$}{Weights of representative days}
\nomenclature[P]{$\mathcal{P}$}{Ambiguity set}
\nomenclature[P]{$c^{EB}$}{Annualized electric boiler cost}
\nomenclature[P]{$c^{conv}$}{Annualized converter cost}
\nomenclature[P]{$c^{comp}$}{Annualized compressor cost}
\nomenclature[P]{$c^{tank}$}{Annualized hydrogen tank cost}
\nomenclature[P]{$M$}{Arbitrarily large number}
\nomenclature[P]{$m^{tank,max}$}{Maximal tank capacity}
\nomenclature[P]{$\mathbf{P}$}{Power production of various corner points of CHP plant}
\nomenclature[P]{$\mathbf{C}$}{Fuel costs of various corner points of CHP plant}
\nomenclature[P]{$\mathbf{P_{H2}}$}{Hydrogen power of various corner points of P2HH}
\nomenclature[P]{$\mathbf{T}$}{Temperature of various corner points of P2HH}
\nomenclature[P]{$\mathbf{P_{Heat}}$}{Heat power of various corner points of P2HH}
\nomenclature[P]{$\mathbf{Q}$}{Heat production of various corner points of CHP plant}
\nomenclature[P]{$\mathbf{m}$}{Wind forecast mean}
\nomenclature[P]{$\eta^{conv}$}{Converter efficiency}
\nomenclature[P]{$\eta^{comp}$}{Compressor efficiency}
\nomenclature[P]{$\eta^{EB}$}{Power to heat conversion efficiency of electric boiler}
\nomenclature[P]{$\mathbf{d}^p$}{Electric demand}
\nomenclature[P]{$\mathbf{d}^q$}{Heat demand}
\nomenclature[p]{$T_{min}, T_{max}$}{Minimum/maximum electrolysis cell temperature}
\nomenclature[P]{$C$}{Specific heat capacity of electrolysis cell}
\nomenclature[P]{$R^{eqv}$}{Equivalent thermal resistance of electrolysis cell}
\nomenclature[P]{$T_a$}{Environment temperature}
\nomenclature[P]{$\epsilon$}{Confidence level parameter}
\nomenclature[P]{$c^{H2}$}{Hydrogen price}
\nomenclature[P]{$Z$}{Number of wind power plants}

\nomenclature[V]{$n^{el}$}{Electrolysis cell number in P2HH stack}
\nomenclature[V]{$\mathbf{u}$}{ON/OFF status of CHP plant}
\nomenclature[V]{$\mathbf{u^{su}},\mathbf{u^{sd}}$}{Binary variable indicating start-up/shut-down of CHP plant}
\nomenclature[V]{$\widetilde{\mathbf{x}},\mathbf{x}$}{Assigned weights of corner points in CHP's operation region}
\nomenclature[V]{$\widetilde{\mathbf{y}},\mathbf{y}$}{Assigned weights of corner points in P2HH's operation region}
\nomenclature[V]{$P^{EB}$}{Electric boiler capacity}
\nomenclature[V]{$P^{conv}$}{Converter capacity}
\nomenclature[V]{$m^{comp}$}{Compressor capacity: maximum flow rate}
\nomenclature[V]{$m^{tank}$}{Tank capacity}
\nomenclature[V]{$\widetilde{p}^{P2HH}, p^{P2HH}$}{Injected power into P2HH}
\nomenclature[V]{$\widetilde{p}^{p2hh}, p^{p2hh}$}{Injected power into single electrolysis cell}
\nomenclature[V]{$\widetilde{h}^{P2HH}, h^{P2HH}$}{Hydrogen production power from P2HH}
\nomenclature[V]{$\widetilde{h}^{p2hh}, h^{p2hh}$}{Hydrogen production power from single electrolysis cell}
\nomenclature[V]{$\widetilde{q}^{P2HH}, q^{P2HH}$}{Heat release from P2HH}
\nomenclature[V]{$\widetilde{q}^{p2hh}, q^{p2hh}$}{Heat release from single electrolysis cell}
\nomenclature[V]{$\widetilde{q}^{EXC}, q^{EXC}$}{Heat exchange with district heating network from P2HH}
\nomenclature[V]{$\widetilde{q}^{exc}, q^{exc}$}{Heat exchange with district heating network from single electrolysis cell}
\nomenclature[V]{$q^{Dissipation}$}{Heat dissipation to the environment from P2HH}
\nomenclature[V]{$U_{tn}$}{Thermal-neutral voltage}
\nomenclature[V]{$U_{rev}$}{Reversible voltage}
\nomenclature[V]{$U_{ohm}$}{Ohmic over-potential}
\nomenclature[V]{$U_{act}$}{Activation over-potential}
\nomenclature[V]{$i_{cell}$}{Electrolysis cell current density}
\nomenclature[V]{$\widetilde{p}^{trans}, p^{trans}$}{Transmitted power from transmission grid}
\nomenclature[V]{$\widetilde{p}^{EB}, p^{EB}$}{Electric boiler consumed power}
\nomenclature[V]{$\widetilde{q}^{EB}, q^{EB}$}{Electric boiler released heat power}
\nomenclature[V]{$\widetilde{n}^{H2}, n^{H2}$}{Produced hydrogen mass}
\nomenclature[V]{$\widetilde{T}, T$}{Electrolysis cell temperature}
\nomenclature[V]{$\widetilde{m}^{H2}, m^{H2}$}{Stored hydrogen mass}
\nomenclature[V]{$z$}{Auxiliary binary variables}
\nomenclature[V]{$e$}{Auxiliary continuous variables}
\nomenclature[V]{$\alpha, \beta, \rho, etc$}{Participation factors}

%% file: 1.intro.tex
\section{Introduction}\label{sec:intro}
As a response to clean energy targets, distributed wind generators have been emerging in distribution networks. However, their uncertainty and variability challenge grid operations, thus limit their continuing rapid development and grid integration. One of the main concerns for distribution networks with a high wind energy penetration level is inverse power flow \cite{Marra2014}, i.e., power flow from low-voltage networks to high-voltage networks. This inverse power flow, typically with a longer delivery path, raises considerable security issues and incurs high power losses \cite{Marra2014, Li2019}.  

Inflexible combined heat and power (CHP) plants have been recognized as an important source of the inverse power flow \cite{Li2019, Chen2018}. For some provinces in China, CHP plants supply nearly 70\% of heat demands \cite{Ge2020}, which inevitably, generate a large amount of electricity at the same time. Especially in winters when energy systems face high heat loads, CHP plants are operated in a heat-demand-driven mode, which implies that CHP plants are operated to follow the high heat demand, thus lose operational flexibility. The inflexible high electricity output from CHP plants, together with uncertain and variable wind power, represents the two significant sources of excess power in distribution networks, challenging the system operation and reducing the energy efficiency.

Under this concern, additional flexibility resources have been introduced into the distribution networks to utilize the excess power, such as electrical energy storage facilities \cite{LUO2015511}, thermal storage units \cite{FINCK2018409} and Power-to-X infrastructures \cite{BLOESS20181611}. In this work, we focus on investigating the couping with power to hydrogen and heat (P2HH) infrastructures (specifically on electrolysers) and electric boilers to build an integrated energy system (IES) to provide flexibility to incorporate the electric and heat loads and address the long-term energy sufficiency. Electrical storage and heat storage are not within the scope of this study based on the following reasons. But it is noteworthy that these two technologies can be incorporated into our problem with slight modification of the model.

\begin{enumerate}
    \item Electrical energy storage though can absorb excess electricity, does not relieve CHP plants from their tight operations, i.e., CHP plants still have to work under the heat-demand-driven mode.
    \item Large-scale electrical energy storage is costly and relies on governmental subsidies to be profitable for power system applications \cite{KHALID2018764}.
    \item The main concern for the integrated energy system is lack of heat supply, which leads CHP plants to operate in the heat-demand-driven mode. Thermal storage units themselves do not supply heat, but rely on power to heat facilities to convert excess power to heat to charge the thermal storage. By introducing P2HH facilities and electric boilers into the system, the excess power can be absorbed. Equivalently important, the CHP plants are released from the heat-demand-driven mode as part of the heat load is supplied by power to heat infrastructures. These two factors combined can significantly reduce the excess power.
    \item Thermal storage units typically suffer from low round-trip efficiency. 
\end{enumerate}

In an overview, this work investigates P2HH facility and electric boiler's roles for flexibility provision in an integrated energy system. The following research questions are addressed.
\begin{enumerate}
    \item How can we determine the optimal sizing of P2HH infrastructure and electric boiler in an integrated energy system, and how do they perform in terms of system cost decrease, CHP plant flexibility expansion and excess power reduction?
    \item How can wind power uncertainties be characterized in an integrated energy system model while not knowing its exact probability distribution? How can uncertainties from power systems propagate to gas and heat systems?
    \item How is profit distributed among various stakeholders in the integrated energy system, and how can we incentivize investments in flexibility resources?
\end{enumerate}  

This study offers an optimal planning decision of the flexibility resources for the integrated energy system considering minimized overall system cost. It is noteworthy that this investment decision falls in individual investors under the market context, who aim to maximize their own profits. However, this  centralized integrated energy system model can still be useful as it can serve as an ideal benchmark to provide insights for policy-makers \cite{Pourahmadi2019}. They can make proper regulations to incentivize potential investors to make investment decisions according to the planning result from this centralized model. In section \ref{rev}, the profit distribution among various stakeholders in the integrated energy system is analyzed. Regulations to incentivize investments of these flexibility resources are also discussed. In the following section, a literature review of most relevant studies is presented to further disclose the contributions of our study.

Existing studies \cite{Chen2018, Chen2015, Hedegaard2012, Meibom2007} have looked at using power to heat infrastructures (e.g., electric boilers, heat pumps, combined with thermal storage) to release CHP plants from the binding heat and power production constraint, thus facilitating wind power integration. An equally important scheme to utilize excess power is turning power to hydrogen  (P2H) via electrolysers. Wang et al. \cite{Wang2019} has looked at the role of power to hydrogen and electric boilers in expanding CHP plant's operational flexibility. An integrated power, heat and hydrogen optimization model was developed to simulate the system operation. The CHP plant's operational flexibility expansion was visualized. However, heat recovery from the electrolyser is not considered in this study, which could account for around 30\% of injected power \cite{Li2019, Ursua2012}. A fixed electricity to hydrogen conversion ratio of the electrolyser was adopted, which neglects its varying temperature-hydrogen-heat (T-H-H) relation that reflects P2HH's non-linear hydrogen and heat production at various temperature. A detailed description of this concept will be seen in section \ref{sec:ies}.

References \cite{Li2019, Ge2020, Fu2019} have proposed operational models for P2HH infrastructures in various integrated energy systems considering heat recovery and T-H-H model. The studies demonstrated the benefits of introducing P2HH in terms of expanded CHP plant flexibility, reduced overall costs and improved local power balancing. However, their models are based on pre-known P2HH sizes. Uncertainties are not addressed in \cite{Ge2020,Fu2019}. A single-level robust optimization (RO) model was adopted in \cite{Li2019} to handle wind and solar power generation uncertainties, which generally provides over-conservative planning results. A joint sizing and operational model for P2HH and other facilities in a multi-carrier energy system was proposed in \cite{Pan2020}. The study applied a two-stage robust optimization (also referred as adaptive robust optimization, ARO) model to account for generation and load uncertainties. Likewise, this model generally gives over-conservative results. Moreover, the conservativeness of the model cannot be easily adjusted. A full comparison between these studies and this paper are provided in Table \ref{tab:modelCompare} to disclose the contributions of this work.

\begin{table*}[tbp]
 \caption{Comparison between developed model with models in literature}\label{tab:modelCompare}
    \centering
             \begin{threeparttable}
    \begin{tabular}{ccccccc}
    \hline
      \textbf{Ref.}  & \textbf{System spec.} &\textbf{\makecell{P2HH \\ sizing}}& \textbf{\makecell{Heat\\ recovery}} & \textbf{\makecell{Electrolyser\\ modeling}} & \textbf{\makecell{Uncertainty\\ handling}} & \textbf{Math. model}\\ \hline
      \cite{Li2019} &CHP,P2HH&No&Yes&T-H-H&RO&MIQCP\\
      \cite{Ge2020}&CHP,P2HH,EB&No&Yes&T-H-H&Deterministic&MILP\\
 \cite{Wang2019}&CHP,P2HH,EB&No&No&Fixed ratios&Deterministic&MILP\\
        \cite{Fu2019}&CHP,P2HH,HCNG&No&Yes&T-H-H&Deterministic&MILP\\
         \cite{Pan2020}&\makecell{GT,P2HH,EB,AC,\\TT,BT,SHS}&Yes&Yes&T-H-H&ARO&\tnote{*}\\
        Paper & CHP,P2HH,EB&Yes & Yes & T-H-H & DRCC&MISOCCQP\\
         \hline
    \end{tabular}
    \begin{tablenotes}
    \footnotesize
    \item[*] Two-stage RO, solved with C\&CG 
    \end{tablenotes}
    \end{threeparttable}
\end{table*}

As indicated in above literature, uncertainty handling has been an important consideration in energy system models. Uncertainty sources in energy systems include power generation, loads, market prices and etc. Scenario-based stochastic programming and robust optimization are the two most common approaches to address uncertainties. Scenario-based stochastic programming approach requires pre-known probability distribution of the uncertainty sources in order to create the scenarios. Moreover, a large number of scenarios is generally necessary to characterize the probability distribution, which significantly undermines computational tractability. Scenario reduction techniques such as K-means clustering algorithm bring back computational tractability, but at the risk of an inappropriate representation of the uncertainties \cite{Pourahmadi2019}. Robust optimization aims to identify the worst uncertainty realization in the uncertainty set and make decisions accordingly. It has found its application in previous studies \cite{Li2019, Pan2020}. However, robust optimization usually leads to over-conservative results. More importantly, its conservativeness cannot be easily adjusted.

By leveraging the advantages of stochastic programming and robust optimization, distributionally robust optimization (DRO) has gain increasing popularity in recent integrated energy system planning and operations studies \cite{He2019IES1, Shahidehpour2021IES2, Shui2019ies4, Zhang2020IES3, Zhang2019ies5, Zhou2019ies6, He2020ies7}. It brings important characteristics such as adjustable conservativeness level and not requiring the uncertainty source's true probability distribution \cite{Shui2019ies4, Zhang2020IES3}. An ambiguity set which contains a group of probability distributions is constructed to characterize the uncertainties. There are different paradigms to construct the ambiguity set in the literature. The studies \cite{Zhou2019ies6, He2020ies7, Ratha2020, Pourahmadi2019} constructed their ambiguity sets based on the first and second order moment information, while in \cite{He2019IES1, Zhang2019ies5} more general moment information was adopted to achieve less conservative results. The authors in \cite{Shui2019ies4} built the ambiguity set based on 1-norm and inf-norm constraints, which allows a simpler solution method. The study \cite{Zhang2020IES3} proposed the use of confidence bands to construct their ambiguity set to incorporate the shape information of uncertaintuy distribution. In \cite{Shahidehpour2021IES2}, the authors proposed a strengthened ambiguity set based on both moment and Wasserstein metric information to characterize the ambiguity set more accurately. Despite the interesting properties of other methods, this study followed the most common approach and constructed the ambiguity set based on  first and second order moment information considering its easier formulation combined with linear decision rules and computational tractability properties \cite{MohajerinEsfahani2018}.  To the best of our knowledge, this modeling technique has not yet been applied for IES planning study incorporating P2HH.

Based on the above literature review, the main contributions of this work fall in the following aspects.
\begin{enumerate}
    \item The sizing of P2HH facility and electric boiler, as well as system operations are jointly optimized for an integrated electricity and heat energy system considering minimized overall system cost.
    \item A distributionally robust chance-constrained integrated energy system model is developed to characterize wind power uncertainties for the integrated energy system. This allows adjusting model conservativeness by selecting different confidence levels in the chance constraints.
    \item Linear decision rules are applied to analytically express real-time control actions when uncertainties are exposed, which allows the propagation of uncertainties originated from power systems to heat and gas systems.
\end{enumerate}

The following part of this study is structured as below. Section \ref{sec:ies} introduces the integrated energy system and derives the T-H-H relation for the P2HH infrastructure. Section \ref{sec:model} details the distributionally robust chance-constrained planing model and linear decision rules that are applied to represent recourse actions. Section \ref{sec:case} presents a case study. Relevant results are shown in section \ref{sec:results}. Section \ref{sec:conclusions} draws conclusions and discusses future work.

%% file: 2.components.tex
\section{Integrated Energy System}\label{sec:ies}
\subsection{Integrated Energy System Structure}
\begin{figure}[tbp]
    \centering
    \includegraphics[width=\linewidth]{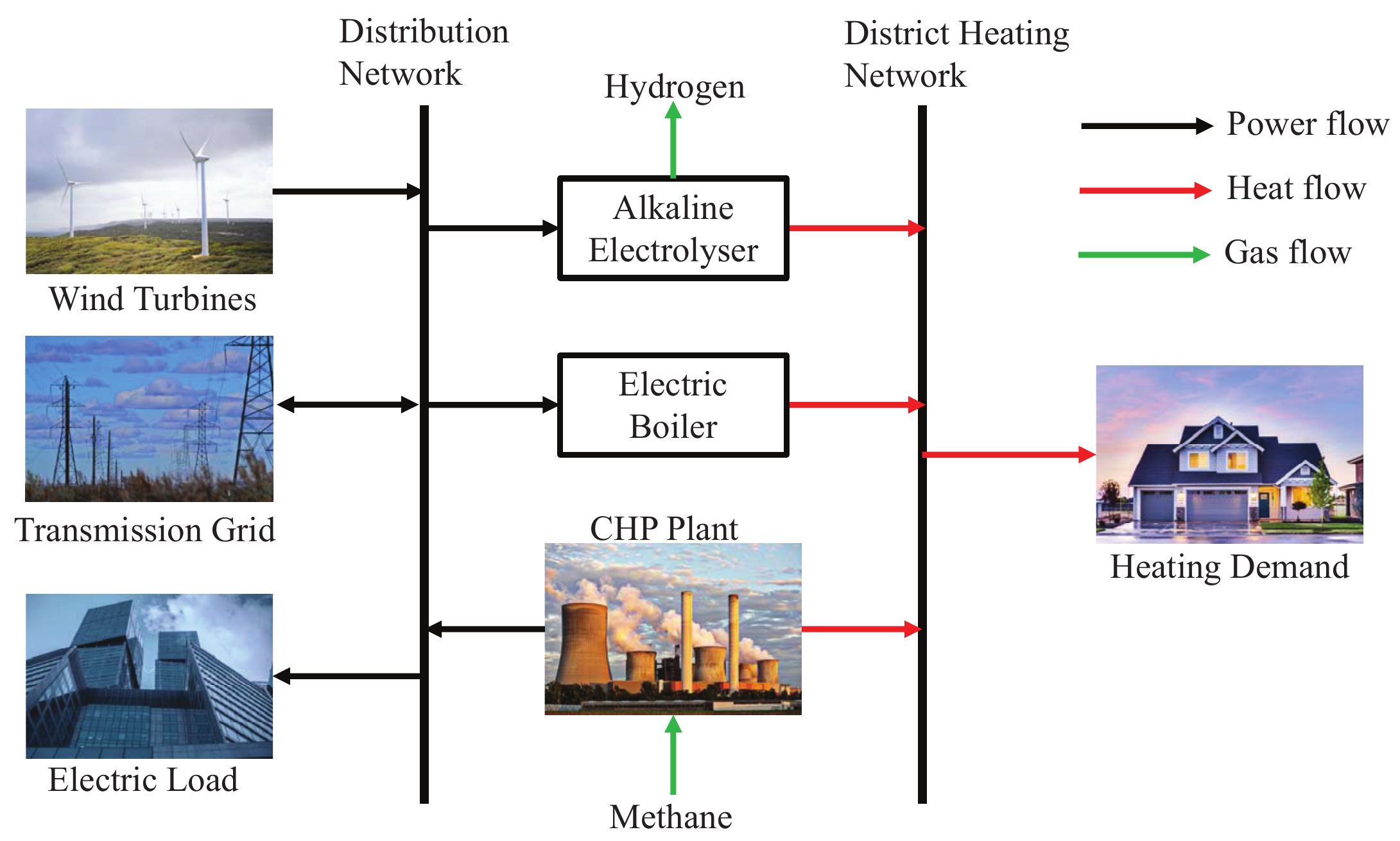}
    \caption{Integrated energy system}
    \label{fig:ies}
\end{figure}

The investigated integrated energy system is presented in Figure \ref{fig:ies}, which consists of existing wind turbines, CHP plant, electric and heating loads and potentially installed power to heat infrastructures: P2HH and electric boiler. Bi-directional power exchange with the transmission grid is also considered. Various types of energy flows are specified, including power flow, heat flow and gas flow (hydrogen and methane). The wind turbines and CHP plant inject power to the distribution network, while P2HH, electric boilers and loads consume power. The distribution network can import or export electricity from the transmission grid for local balancing. Produced heat from the P2HH infrastructure, electric boiler and CHP plant is injected to the district heating network to satisfy the heating demand. Similar to \cite{Pourahmadi2019, Ratha2020}, the electric and heating loads are assumed inelastic to price. It is noteworthy that an alkaline electrolyser (AEC) is selected for this P2HH application due to the fact that it has been commercialized in Mega-Watt levels \cite{Li2019, Pan2020}, which so far is not the case for its counterparts: polymer electrolyte membrane electrolyser (PEMEC) and solid oxide electrolyser (SOEC).

\subsection{P2HH Modeling}
In this section, we focus on the derivation of the T-H-H model that was originally proposed in \cite{Li2019} and applied in \cite{Ge2020, Fu2019, Pan2020} to model the non-linear temperature, hydrogen and heat production relation of P2HH. It is noted in the previous part that heat release accounts for approximately 30\% of injected power into the P2HH. By recovering this part of low-grade energy (60-80\textdegree{C} \cite{Dincer2014}) for district heating use, the overall efficiency of P2HH can be significantly increased. Therefore, it is important to have an accurate model to account for the hydrogen and heat production in the P2HH. In references \cite{Wang2019, Hou2017}, a fixed power to hydrogen conversion factor for the P2HH infrastructure has been assumed, which can be easily integrated into optimization models. However, at varying temperature and varying current density, this assumption fails as this conversion factor can be changing significantly, which will be shown in later this chapter. In the following sections, \ref{sec:P2HHstructure} explains the P2HH structure, \ref{sec:THH} derives the T-H-H relation.

\subsubsection{P2HH Structure}\label{sec:P2HHstructure}

\begin{figure}[tbp]
    \centering
    \includegraphics[width=\linewidth]{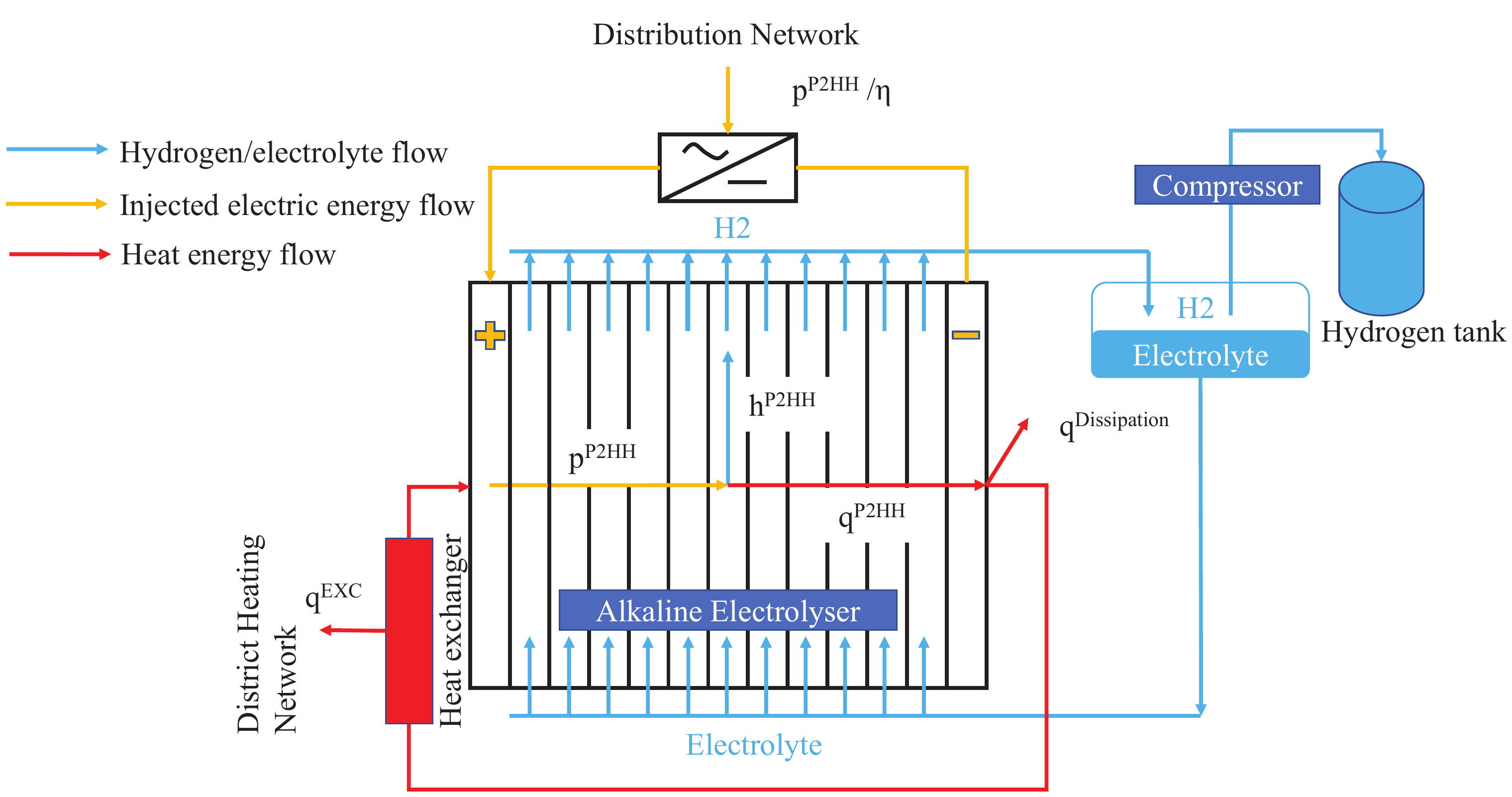}
    \caption{P2HH Structure}
    \label{fig:P2HH}
\end{figure}

A schematic configuration of P2HH infrastructure is presented in Figure \ref{fig:P2HH}, which consists of an AC/DC converter, alkaline electrolytic cells, a heat exchanger which collects heat for district heating network use, a compressor and a hydrogen tank, which stores hydrogen on a daily basis. The produced hydrogen can be used in the transport sector or industry. Various energy flows are specified in the figure, which include hydrogen energy flow, injected electric energy flow and heat energy flow. It is worth noting that we use capital abbreviation P2HH to refer to the electrolyser stack while we use the lowercase $p2hh$ to refer to each electrolytic cell. The number of electrolytic cells in the stack is an integer decision variable to be determined in our model, which links single cells and the stack.

The working process of P2HH is explained as below. The injected power $p^{P2HH}$ is split into power for hydrogen $h^{P2HH}$ and heat $q^{P2HH}$ production. The amount of produced hydrogen is directly proportional to hydrogen production power $h^{P2HH}$. Hydrogen is separated from the electrolyte and collected in the hydrogen tank. Electrolyte in the stack is accordingly supplemented. The other part, i.e., heat power $q^{P2HH}$, deducted from dissipation to the environment $q^{Dissipation}$, is collected by a close-loop water circulation system to supply district heating demand through a heat exchanger. It is noteworthy that this recycled heat $q^{EXC}$ is not necessarily equivalent to the available heat $q^{P2HH}-q^{Dissipation}$. Instead, by manipulating $q^{EXC}$, electrolysis stack temperature can be adjusted in the model.

\subsubsection{T-H-H Relation}\label{sec:THH}
In this section, we derive the T-H-H relation for a single P2HH cell, which will be integrated into the model by assuming all the electrolytic cells are working under the same condition \cite{Li2019}. The reason for deriving this T-H-H relation is that at different temperature and different current density, power to hydrogen and heat ratios can be varying significantly, which in return change the cell temperature according to temperature evolution in the cell. The derivation is shown as below.

\begin{subequations}\label{eq:THH}
\begin{equation}\label{eq:powerDensity}
    p^{p2hh} = U_{cell} i_{cell}
\end{equation}
\begin{equation}\label{eq:cellVoltage}
\begin{split}
    U_{cell}\left(i_{cell}, T\right) = U_{rev}(T)+U_{ohm}\left(i_{cell}, T\right)\\+U_{act}\left(i_{cell}, T\right)
\end{split}
\end{equation}
\begin{equation}\label{eq:hydrogenpowerDensity}
    h^{p2hh} = U_{tn} i_{cell}
\end{equation}
\begin{equation}\label{eq:thermalNeutralvoltage}
    U_{tn} = U_{tn}(T)
\end{equation}
\begin{equation}\label{eq:heatpowerDensity}
    q^{p2hh}= p^{p2hh}-h^{p2hh}= (U_{cell}-U_{tn}) i_{cell}
\end{equation}
\begin{equation}\label{T-H-H}
    q^{p2hh}= \frac{ h^{p2hh}}{U_{tn}(T)} \Bigg(U_{cell}\bigg(\frac{ h^{p2hh}}{U_{tn}(T)}, T \bigg) - U_{tn}(T)\Bigg)
\end{equation}
\end{subequations}

Injected power density of the P2HH infrastructure $p^{p2hh}$ is stated as (\ref{eq:powerDensity}), where $i_{cell}$ refers to the cell current density and $U_{cell}$ refers to the cell voltage. The polarization of the electrolysis cell is stated as Eq. (\ref{eq:cellVoltage}), which imposes that the cell voltage is composed of reversible voltage $U_{rev}$, over-potentials from ohmic loss $U_{ohm}$ and activation $U_{act}$. Concentration over-potential from mass transport limitation is neglected as it is much smaller compared to $U_{ohm}$ and $U_{act}$, especially for alkaline electrolysers \cite{Ursua2012, Koponen2015}. According to \cite{Li2019, Pan2020}, hydrogen production power density can be stated as (\ref{eq:hydrogenpowerDensity}) and (\ref{eq:thermalNeutralvoltage}). Combining (\ref{eq:powerDensity}) to (\ref{eq:thermalNeutralvoltage}), the released heat can be expressed as (\ref{eq:heatpowerDensity}), being the difference between injected power and power used for hydrogen production. By eliminating current density $i_{cell}$ in (\ref{eq:heatpowerDensity}), the T-H-H relation is presented as (\ref{T-H-H}), which establishes the non-linear relation between temperature, hydrogen production and heat release for the P2HH facility. It is noteworthy that eliminating $i_{cell}$ does not affect the model accuracy, as $i_{cell}$ is determined when temperature and hydrogen power are known, shown as (\ref{eq:hydrogenpowerDensity})-(\ref{eq:thermalNeutralvoltage}).
Referring to \cite{Koponen2015, Roy2006, Gilliam2007}, detailed derivation of cell voltage and thermal-neutral voltage and data are available on the online open-source repository \cite{SenTHH}.

\begin{figure}[tbp]
    \centering
    \includegraphics[width=.9\linewidth]{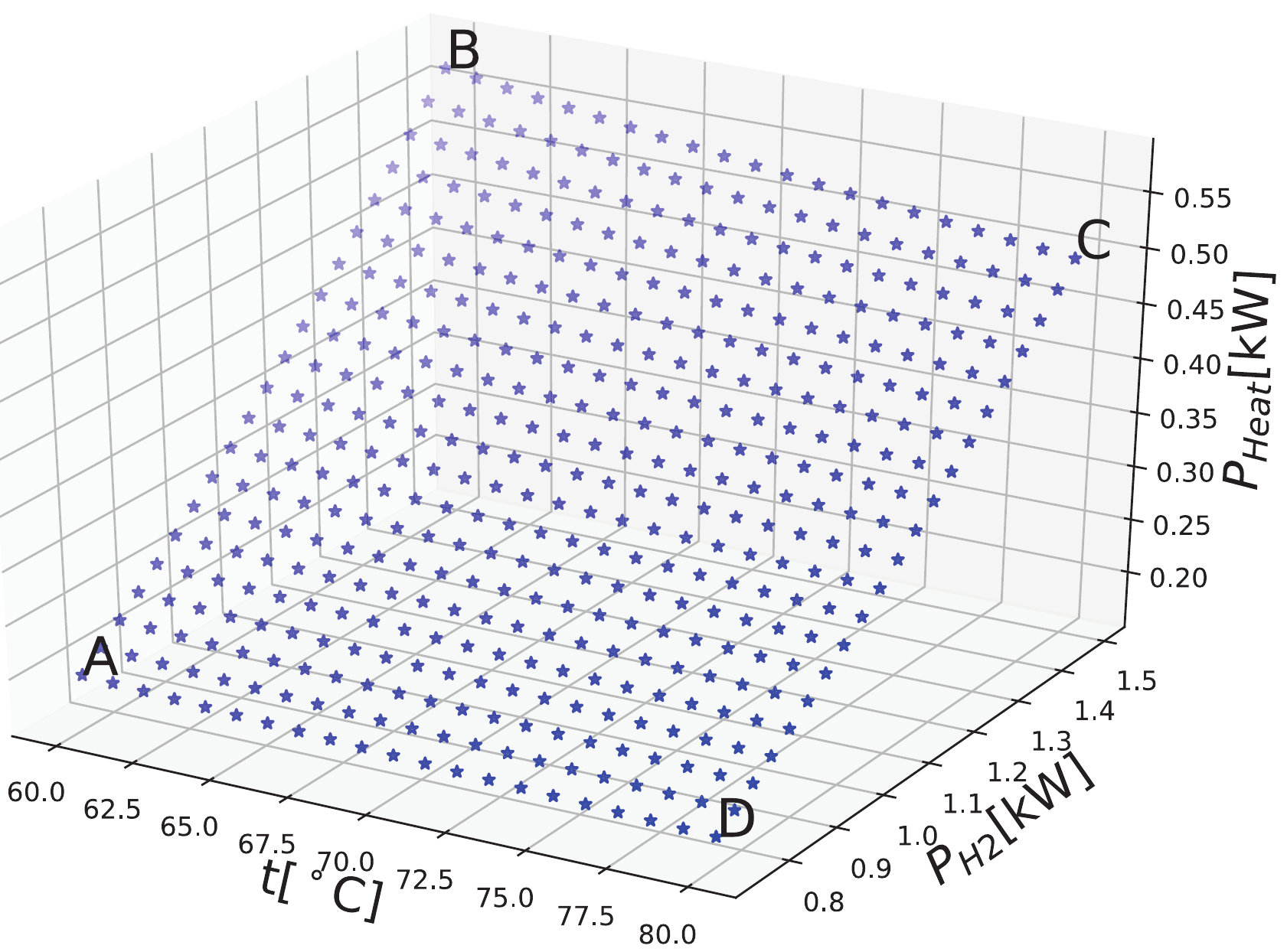}
    \caption{T-H-H relation for P2HH}
    \label{fig:THH}
\end{figure}

The obtained T-H-H relation for the alkaline electrolyser is illustrated in Figure \ref{fig:THH}, which presents an approximately affine surface. Each point on the surface corresponds to a different temperature and current density combination. Power instead of power density is used assuming a cell area of 0.25$\mathrm{m^2}$. The boundaries reflect temperature limits and current density limits. Specifically, AB and CD reflects lower and upper temperature limits (60-80\textdegree{C}), AD and BC reflects lower and upper current density limits (0.2-0.4$\mathrm{Acm^{-2}}$). These limits are imposed from \cite{Dincer2014}. The convex hull formulated by the boundary points is adopted to approximate this non-linear surface in order for this T-H-H relation model to be easily into the following optimization model.

\begin{figure}[tbp]
    \centering
    \includegraphics[width = 0.9\linewidth]{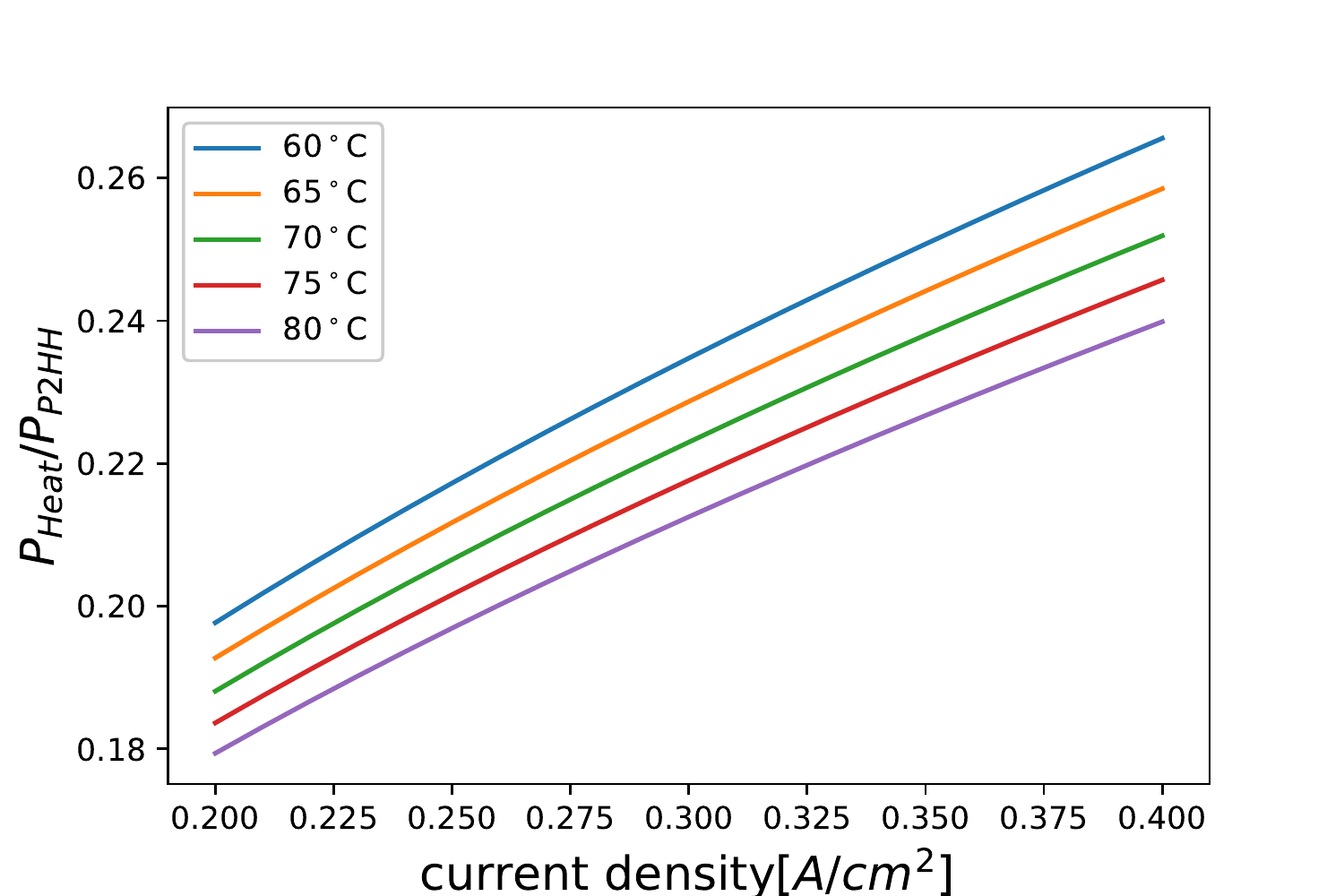}
    \caption{Power to heat ratios at various temperature and current density}
    \label{fig:heatratio}
\end{figure}

\begin{figure}[tbp]
    \centering
    \includegraphics[width = 0.9\linewidth]{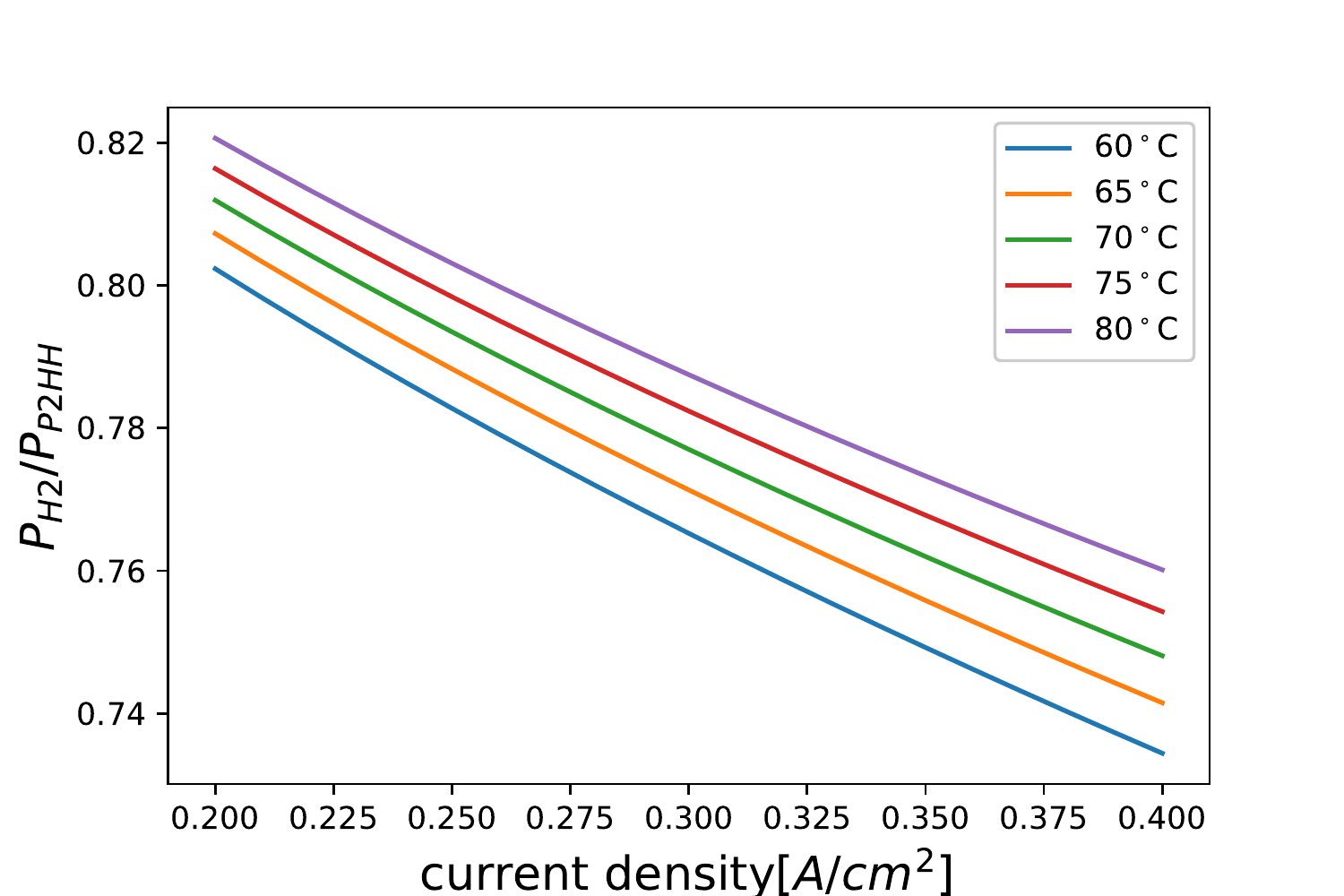}
    \caption{Power to hydrogen ratios at various temperature and current density}
    \label{fig:h2ratio}
\end{figure}

Power to heat and hydrogen ratios are illustrated in Figure \ref{fig:heatratio} and Figure \ref{fig:h2ratio} respectively. At higher current densities, thermal-neutral voltage remains the same, while cell voltage increases significantly. Therefore, power to hydrogen ratio decreases, while power to heat ratio increases as a higher proportion of power is converted to heat. At increasing temperatures, thermal-neutral voltage increases slowly. However, cell voltage drops as higher temperature facilitates electrolysis reaction. Therefore, an increase in power to hydrogen ratio and a decrease in power to heat ratio are seen. Combining the influence of temperature and current density, power to heat ratio ranges from 0.18 to 0.26 (44.4\% difference), power to hydrogen ratio ranges from 0.74 to 0.82 (10.8\% difference), which underlines the importance of applying a more accurate T-H-H relation rather than assuming fixed power to heat and power to hydrogen ratios.

%% file: 3.MathModel.tex
\section{Problem Formulation}\label{sec:model}
\subsection{Addressing Uncertainties}
As previously introduced, a distributionally robust chance-constrained planning model is applied to address short-term wind power uncertainties while assuming inelastic (i.e., deterministic) electric and heating demands. These short-term wind power uncertainties are modeled via ambiguity sets, which are a family of distributions concerning wind power forecast errors having the same first- and second-order moment information, i.e., mean and covariance. It is assumed that exact values for the first- and second-order moments can be estimated from historical data \cite{Pourahmadi2019}.

We note $Z$ as the number of uncertainty sources, i.e., wind power generators. The wind power production in the short term is modeled as $\mathbf{m}_{r,t} + \boldsymbol{\omega}_{r,t}$, where $\mathbf{m}_{r,t}\in\mathbb{R}^{Z}$ refers to the wind power forecast vector and $\boldsymbol{\omega}_{r,t} \in \mathbb{R}^{Z}$ refers to the uncertain wind power forecast error vector which follows some unknown distribution in the ambiguity set $\mathcal{P}_{r,t}$. Without loss of generality, the first-order moment, i.e., mean of wind power forecast errors, similar to \cite{Pourahmadi2019, Ratha2020}, is assumed to be $\mathbf{0}$, i.e., $\boldsymbol{\mu}_{r,t} = \mathbf{0}$. Using this zero-mean assumption, we note the ambiguity set $\mathcal{P}_{r,t}$ as (\ref{eq:ambiguitySet}) for each representative day and each hour, which includes a family of distributions $D$ in $\Psi(\mathbb{R}^Z)$ that have the same mean $\boldsymbol{\mu}_{r,t}$ and covariance $\boldsymbol{\Sigma}_{r,t}$. $\mathbb{E}$ refers to the expectation operator wherein the uncertain parameter $\boldsymbol{\omega}$ follows the distribution $D$.
\begin{equation}\label{eq:ambiguitySet}
    \mathcal{P}_{r,t}=\left\{D \in \Psi \left(\mathbb{R}^{Z}\right): \mathbb{E}^{D}(\boldsymbol{\omega})=\boldsymbol{\mu}_{r, t}, \mathbb{E}^{D}\left( \boldsymbol{\omega} \boldsymbol{\omega}^{\top}\right)=\boldsymbol{\Sigma_{r, t}}\right\}
\end{equation}

The underlying philosophy behind distributionally robust chance-constrained planning is that we would like to identify the worst distribution in this ambiguity set $\mathcal{P}_{r,t}$ that $\boldsymbol{\omega}_{r,t}$ follows and make decisions under this worst distribution, hence is consistent with the idea of \textit{distributionally robust} planning. Using the above ambiguity set, the distributionally robust chance-constrained model is formulated in the following sections.

\subsection{Mathematical Model}
The objective function (\ref{eq:obj}) seeks to minimize the total system cost, taking into account investment and O\&M costs of the electrolyser, AC/DC converter, compressor, hydrogen tank and electric boiler, and system operational costs, where the set $\Theta$ includes decision variables in the planning stage, i.e., $\{n^{el}, P^{conv},\\ m^{comp}, m^{tank}, P^{EB}, \mathbf{u^{su}, u^{sd},\mathbf{u}}\}$. The first five terms, i.e., $c^{el} n^{el} +c^{conv}P^{conv} + c^{comp}m^{comp}+ c^{tank} m^{tank} + c^{EB}P^{EB}$, are the annualized investment and O\&M costs of the electrolyser, AC/DC converter, compressor, hydrogen tank and electric boiler, whereas the next term represents the start-up and shut-down costs of the CHP plant for a target year, with $r$ and $k_r$ referring to a set of representative days and their corresponding weights for the year.
\begin{equation}\label{eq:obj}
\begin{split}
  \underset{\Theta}{\min.}  \Bigg\{ c^{el} n^{el} + c^{conv}P^{conv} + c^{comp}m^{comp} +c^{tank} m^{tank}\\ + c^{EB}P^{EB} + \sum_r k_r \sum_t \bigg[c^{su} u_{r,t}^{su} + c^{sd} u_{r,t}^{sd} \bigg ] \\+ \underset{D \in \mathcal{P}_{rt}}{\max.}  \underset{\Theta^{rec}}{\min.} \sum_r k_r \sum_t  \mathbb{E}^{D}  \bigg[\mathbf{C}^{\top} \mathbf{\widetilde{x}}_{r,t}(\boldsymbol{\omega}_{r,t})  \\ + c^{trans} (\widetilde{p}_{r,t}^{trans}(\boldsymbol{\omega}_{r,t}))^2 -c^{H2} \widetilde{n}_{r,t}^{H2}(\boldsymbol{\omega}_{r,t})\bigg]\bigg \}   
\end{split}
\end{equation}

The last term includes the CHP operation cost, electricity exchange cost with the transmission grid and hydrogen sale profit respectively. The electricity exchange cost is a quadratic term, penalizing both  importing and exporting electricity from the transmission grid. It is noteworthy that this last term takes a max-min form, aiming to identify the worst forecast error distribution in the ambiguity set (thus distributionally robust) and optimize recourse actions based on the worst distribution, given the investment decisions and day-ahead CHP plant ON/OFF decisions. Using linear decision rules, this overall three-stage (i.e., min-max-min) distributionally robust problem boils down to a single-level minimization problem and can be readily solved with existing solvers. Details will be covered in later this section.
\begin{equation}\label{eq:facility}
    n^{el} \in \mathbb{N}, [P^{conv}, m^{comp}, P^{EB}] \geq 0, 0 \leq m^{tank} \leq m^{tank, max}
\end{equation}

Constraints (\ref{eq:facility}) is imposed on the investment decision variables, where the number of electrolysis cells in the electrolyser stack $n^{el}$ is constrained as an integer variable. An upper limit is set to the tank size $m^{tank}$ for practical consideration of a maximum hydrogen export on a daily basis. 
\begin{subequations}\label{eq:CHP}
\begin{equation}\label{eq:integrity}
     \mathbf{u^{su}}\in \mathbb{B}^{|r|\times|t|}, \mathbf{u^{sd}} \in \mathbb{B}^{|r|\times|t|},
    \mathbf{u} \in \mathbb{B}^{|r|\times|t|}
\end{equation}
\begin{equation}\label{eq:startup0}
     u_{r,t} - u_{r,t}^{su} \leq 0, \forall r, t= 1
\end{equation}
\begin{equation}\label{eq:startup}
    -u_{r,t-1} + u_{r,t} - u_{r,t}^{su} \leq 0, \forall r, t \geq 2
\end{equation}
\begin{equation}\label{eq:shutdown}
    u_{r,t-1} - u_{r,t} - u_{r,t}^{sd} \leq 0, \forall r, t \geq 2
\end{equation}
\begin{equation}\label{eq:min_on0}
      u_{r,t} - u_{r,\tau} \leq 0, \forall \tau \in \{t+1,...,\min(|t|,\bar{v}+t-1)\}, \forall r, t = 1
\end{equation}
\begin{equation}\label{eq:min_on}
\begin{split}
     -u_{r,t-1} + u_{r,t} - u_{r,\tau} \leq 0, \forall \tau \in \{t+1,...,\min(|t|,\bar{v}+t-1)\}, \\ \forall r, 2 \leq t \leq |t|-1
\end{split}
\end{equation}
\begin{equation}\label{eq:min_off}
\begin{split}
     u_{r,t-1} - u_{r,t} + u_{r,\tau} \leq 1, \forall \tau \in \{t+1,...,\min(|t|,\underline{v}+t-1)\},\\ \forall r,  2 \leq t \leq |t|-1
\end{split}
\end{equation}
\begin{equation}\label{eq:CHPintegrity}
    \mathbf{1}^{\top} \mathbf{\widetilde{x}}_{r,t}(\boldsymbol{\omega}_{r,t}) = u_{r,t}, \forall r, \forall t
\end{equation}
\begin{equation} \label{eq:chanceXbigger}
 \underset{D \in \mathcal{P}_{rt}}{\min} \mathbb{P}[\mathbf{\widetilde{x}}_{r,t}(\boldsymbol{\omega}_{r,t}) \geq 0] \geq 1 - \epsilon, \forall r, \forall t
\end{equation}
\begin{equation} \label{eq:chanceXless}
 \underset{D \in \mathcal{P}_{rt}}{\min} \mathbb{P}[\mathbf{\widetilde{x}}_{r,t}(\boldsymbol{\omega}_{r,t}) \leq 1] \geq 1 - \epsilon, \forall r, \forall t
\end{equation}
\begin{equation} \label{eq:rampup1}
  \underset{D \in \mathcal{P}_{rt}}{\min} \mathbb{P}[  \mathbf{P}^{\top} \mathbf{\widetilde{x}}_{r,t}(\boldsymbol{\omega}_{r,t})\leq SU ] \geq 1 - \epsilon, \forall r, t = 1
\end{equation}
\begin{equation} \label{eq:rampup2}
\begin{split}
  \underset{D \in \mathcal{P}_{rt}}{\min} \mathbb{P}[  \mathbf{P}^{\top} \mathbf{\widetilde{x}}_{r,t}(\boldsymbol{\omega}_{r,t}) -  \mathbf{P}^{\top} \mathbf{\widetilde{x}}_{r,t-1}(\boldsymbol{\omega}_{r,t-1})  \leq SU (1-u_{r,t-1}) \\ + RU u_{r,t-1}]  \geq 1 - \epsilon, \forall r, t \geq 2
  \end{split}
  \end{equation}
\begin{equation} \label{eq:rampdown2}
\begin{split}
 \underset{D \in \mathcal{P}_{rt}}{\min} \mathbb{P}[ \mathbf{P}^{\top} \mathbf{\widetilde{x}}_{r,t-1}(\boldsymbol{\omega}_{r,t-1}) -  \mathbf{P}^{\top} \mathbf{\widetilde{x}}_{r,t}(\boldsymbol{\omega}_{r,t})  \leq SD (1-u_{r,t}) \\+ RD u_{r,t}]  \geq 1 - \epsilon, \forall r,  t\geq 2
\end{split}
\end{equation}
\end{subequations}

Operational constraints of the CHP plant are shown in (\ref{eq:CHP}), where (\ref{eq:integrity}) defines the start-up, shut-down and ON/OFF status of the CHP plant using sets of binary variables respectively. (\ref{eq:startup0})-(\ref{eq:shutdown}) associate ON/OFF status variables with start-up and shut-down variables which assume the CHP plant to be off at the beginning of each representative days. The minimum up- and down- time limits of the CHP plants are imposed in (\ref{eq:min_on0})-(\ref{eq:min_off}). Constraints (\ref{eq:CHPintegrity})-(\ref{eq:rampdown2}) enforce the real-time operation limits for the CHP plant. (\ref{eq:CHPintegrity}) constrains the CHP plant to be operated within its feasible region if it is on, where $\mathbf{\widetilde{x}}_{r,t}$ is a $\mathbb{R}^4$-valued function with its components summing up to 1, representing weights to each corner points in the operational region (A, B, C, D as in Fig \ref{fig:CHPoperationalRegion}). (\ref{eq:chanceXbigger})-(\ref{eq:rampdown2}) are a set of individual distributionally robust chance constraints, where $\mathbb{P}[\boldsymbol{\cdot}]$ is the probability operator wherein the uncertainty source $\boldsymbol{\omega}_{r,t}$ follows the worst distribution $D$ in the ambiguity set $\mathcal{P}_{r,t}$. This implies that under the worst distribution, the probability of meeting each individual inequality constraints should be greater than or equal to $1-\epsilon$, where $\epsilon$ is a predefined parameter from 0 to 1. This allows adjusting the conservativeness of the developed model by choosing different $\epsilon$. It is worth noting that the worst distribution within the constraints and that in the objective function (\ref{eq:obj}) are not necessarily identical. By adopting linear decision rules and Cantelli's inequality (a one-sided Chebyshev inequality), the distributionally robust chance constraints can be formulated as second-order cone constraints. Specifically, (\ref{eq:chanceXbigger}) and (\ref{eq:chanceXless}) constrain the weights of each corner points in CHP plant's operation region to be within 0 to 1. (\ref{eq:rampup1})-(\ref{eq:rampdown2}) pertain to ramping rates of the CHP plant. 

\begin{subequations}\label{eq:balance}
\begin{equation}\label{eq:powerbalance}
\begin{split}
    \widetilde{p}^{trans}_{r,t}(\boldsymbol{\omega}_{r,t}) + \mathbf{P}^{\top} \mathbf{\widetilde{x}}_{r,t}(\boldsymbol{\omega}_{r,t}) + \mathbf{1}^{\top}(\mathbf{m}_{r,t} + \boldsymbol{\omega}_{r,t}) =    \\ \widetilde{p}^{P2HH}_{r,t}(\boldsymbol{\omega}_{r,t})/\eta^{conv} + \widetilde{n}^{H2}_{r,t} \eta^{comp} +\widetilde{p}^{EB}_{r,t}(\boldsymbol{\omega}_{r,t}) + \mathbf{1}^{\top} \mathbf{d}^{p}_{r,t}, \forall r, \forall t
\end{split}
\end{equation}
\begin{equation}\label{eq:heatbalance}
\mathbf{Q}^{\top} \mathbf{\widetilde{x}}_{r,t}(\boldsymbol{\omega}_{r,t}) +  \widetilde{q}^{EXC}_{r,t}(\boldsymbol{\omega}_{r,t}) +\widetilde{q}^{EB}_{r,t}(\boldsymbol{\omega}_{r,t})= \mathbf{1}^{\top} \mathbf{d}^q_{r,t}, \forall r, \forall t
\end{equation}
\end{subequations}

Similar to (\ref{eq:CHPintegrity}), the electricity and heat balance equality constraints (\ref{eq:balance}) are met regardless of the uncertainties. Specifically, (\ref{eq:powerbalance}) enforces the electricity balance, where the power injection from the transmission grid, CHP plant and distributed wind generators are equal to the power consumption of the P2HH facility, compressor, electric boiler and the inelastic demand. A conversion efficiency parameter $\eta^{conv}$ is applied to account for AC/DC conversion loss. Electricity consumption of the compressor is set to be linearly related to hydrogen production by a constant compression efficiency $\eta^{comp}$ in the unit of $\mathrm{MWh/kg}$, which compresses outlet hydrogen from 30 bar to around 540 bar for truck load \cite{Hou2017}. (\ref{eq:heatbalance}) imposes the heat balance, where the heat output from the CHP plant, P2HH facility and electric boiler equals the inelastic heat demand. Using linear decision rules, the nominal terms and stochastic terms can be separated and the original equality constraint can be replaced as two (nominal and stochastic) corresponding equality constraints. 
\begin{subequations}\label{eq:EB}
\begin{equation}\label{eq:EBbig}
        \underset{D \in \mathcal{P}_{rt}}{\min} \mathbb{P}[ \widetilde{p}^{EB}_{r,t}(\boldsymbol{\omega}_{r,t}) \geq0] \geq 1 - \epsilon, \forall r, \forall t
\end{equation}
\begin{equation}\label{eq:EBless}
\underset{D \in \mathcal{P}_{rt}}{\min} \mathbb{P}[
    \widetilde{p}^{EB}_{r,t}(\boldsymbol{\omega}_{r,t}) \leq P^{EB}] \geq 1 - \epsilon, \forall r, \forall t
\end{equation}
\begin{equation}\label{eq:EBratio}
   \widetilde{q}^{EB}_{r,t}(\boldsymbol{\omega}_{r,t}) =  \widetilde{p}^{EB}_{r,t}(\boldsymbol{\omega}_{r,t}) \eta^{EB}, \forall r, \forall t  
\end{equation}
\end{subequations}

Constraints (\ref{eq:EB}) pertain to operations of the electric boiler, where (\ref{eq:EBbig})(\ref{eq:EBless}) enforce the electric boiler to operate within its designed capacity. (\ref{eq:EBratio}) imposes a linear relation between the power injection and heat output via a fixed power conversion ratio $\eta^{EB}$.

\begin{subequations}\label{eq:P2HH}
\begin{equation}\label{eq:p2hhstackp}
    \widetilde{p}^{P2HH}_{r,t}(\boldsymbol{\omega}_{r,t}) = n^{el} \widetilde{p}^{p2hh}_{r,t}(\boldsymbol{\omega}_{r,t}), \forall r, \forall t
\end{equation}

\begin{equation}\label{eq:p2hhstackh}
    \widetilde{h}^{P2HH}_{r,t}(\boldsymbol{\omega}_{r,t}) = n^{el} \widetilde{h}^{p2hh}_{r,t}(\boldsymbol{\omega}_{r,t}), \forall r, \forall t
\end{equation}
\begin{equation}\label{eq:p2hhexc}
    \widetilde{q}^{EXC}_{r,t}(\boldsymbol{\omega}_{r,t}) = n^{el} \widetilde{q}^{exc}_{r,t}(\boldsymbol{\omega}_{r,t}), \forall r, \forall t
\end{equation}
\begin{equation}\label{eq:excOver0}
     \underset{D \in \mathcal{P}_{rt}}{\min} \mathbb{P}[\widetilde{q}^{exc}_{r,t}(\boldsymbol{\omega}_{r,t}) \geq 0] \geq 1 - \epsilon, \forall r, \forall t
\end{equation}
\begin{equation}\label{eq:elecPower}
   \widetilde{p}^{p2hh}_{r,t}(\boldsymbol{\omega}_{r,t}) = \widetilde{h}^{p2hh}_{r,t}(\boldsymbol{\omega}_{r,t}) +  \widetilde{q}^{p2hh}_{r,t}(\boldsymbol{\omega}_{r,t}), \forall r, \forall t
\end{equation}

Constraint (\ref{eq:P2HH}) pertains to operations of the P2HH facility, where (\ref{eq:p2hhstackp})-(\ref{eq:p2hhexc}) relate the cell operational variables with the electrolyser stack operational variables.  (\ref{eq:excOver0}) imposes that the P2HH facility supplies heat to the district heating network. Power balance of the electrolyser is enforced in (\ref{eq:elecPower}), where the injecting power is converted to hydrogen energy and released heat. 
\begin{equation}\label{eq:h2conversion}
         \widetilde{n}^{H2}_{r,t}(\boldsymbol{\omega}_{r,t}) =       \frac{3.6 \times 10^6}{U_{tn}F} \widetilde{h}^{P2HH}_{r,t}(\boldsymbol{\omega}_{r,t}), \forall r, \forall t
\end{equation}

Using Faraday's law of electrolysis and assuming a constant thermal-neutral voltage in the operational temperature range (0.44\% difference over the temperature range), the produced hydrogen (kg) can be related to the hydrogen production power (MW) as (\ref{eq:h2conversion}), where $F$ is the Faraday's constant. 
\begin{equation}\label{eq:cellIntegrity}
    \mathbf{1}^{\top} \mathbf{\widetilde{y}}_{r,t}(\boldsymbol{\omega}_{r,t}) = 1, \forall r, \forall t
\end{equation}
\begin{equation} \label{eq:chanceybigger}
 \underset{D \in \mathcal{P}_{rt}}{\min} \mathbb{P}[\mathbf{\widetilde{y}}_{r,t}(\boldsymbol{\omega}_{r,t}) \geq 0] \geq 1 - \epsilon, \forall r, \forall t
\end{equation}
\begin{equation} \label{eq:chanceyless}
 \underset{D \in \mathcal{P}_{rt}}{\min} \mathbb{P}[\mathbf{\widetilde{y}}_{r,t}(\boldsymbol{\omega}_{r,t}) \leq 1] \geq 1 - \epsilon, \forall r, \forall t
\end{equation}
\begin{equation}\label{eq:hydrogenPower}
    \mathbf{P_{H2}}^{\top} \mathbf{\widetilde{y}}_{r,t}(\boldsymbol{\omega}_{r,t}) = \widetilde{h}^{p2hh}_{r,t}(\boldsymbol{\omega}_{r,t}), \forall r, \forall t
\end{equation}
\begin{equation}\label{eq:heatingPower}
    \mathbf{P_{Heat}}^{\top} \mathbf{\widetilde{y}}_{r,t}(\boldsymbol{\omega}_{r,t}) = \widetilde{q}^{p2hh}_{r,t}(\boldsymbol{\omega}_{r,t}), \forall r, \forall t
\end{equation}
\begin{equation}\label{eq:temperature0}
    \mathbf{T}^{\top} \mathbf{\widetilde{y}}_{r,t}(\boldsymbol{\omega}_{r,t}) = T_{r,t},\forall r, t = 1
\end{equation}
\begin{equation}\label{eq:temperature}
    \mathbf{T}^{\top} \mathbf{\widetilde{y}}_{r,t}(\boldsymbol{\omega}_{r,t}) = \widetilde{T}_{r,t}(\boldsymbol{\omega}_{r,t}),\forall r,  t \geq 2
\end{equation}

Similar to the CHP plant, the P2HH facility takes a linearized operation region, with $\widetilde{\mathbf{y}}$ which is a $\mathbb{R}^4$-valued function representing weights associated with each corner points in the operation region, hence (\ref{eq:cellIntegrity})-(\ref{eq:chanceyless}). Hydrogen production power, released heat and electrolyte temperature are related to $\widetilde{\mathbf{y}}$ by (\ref{eq:hydrogenPower})-(\ref{eq:temperature}), with $\mathbf{P}_{\mathbf{H2}} \in \mathbb{R}^4$ and $\mathbf{P}_{\mathbf{Heat}} \in \mathbb{R}^4$ denoting hydrogen production power and released heat in the corner points. It is noteworthy that the starting temperature of the P2HH is assumed as a pre-known parameter (e.g., 80\textdegree{C}), which does not take a stochastic term (\ref{eq:temperature0}).
\begin{equation}\label{eq:tempRangeLow}
   \underset{D \in \mathcal{P}_{rt}}{\min} \mathbb{P}[  T_{min} \leq\widetilde{T}_{r,t}(\boldsymbol{\omega}_{r,t})] \geq 1 - \epsilon ,\forall r,  t\geq 2
\end{equation}
\begin{equation}\label{eq:tempRangeHigh}
   \underset{D \in \mathcal{P}_{rt}}{\min} \mathbb{P}[  \widetilde{T}_{r,t}(\boldsymbol{\omega}_{r,t}) \leq  T_{max}] \geq 1 - \epsilon ,\forall r,  t\geq 2
\end{equation}
\begin{equation}\label{eq:tempEvolution0}
\begin{split}
    \widetilde{T}_{r,t+1}(\boldsymbol{\omega}_{r,t+1}) = T_{r,t}  + \frac{1}{C}\Bigg(\widetilde{q}^{p2hh}_{r,t}(\boldsymbol{\omega}_{r,t}) - \widetilde{q}^{exc}_{r,t}(\boldsymbol{\omega}_{r,t}) \\  - \frac{1}{R^{eqv}}(T_{r,t} - T_a)\Bigg),  \forall r, t = 1
\end{split}
\end{equation}
\begin{equation}\label{eq:tempEvolution}
\begin{split}
      \widetilde{T}_{r,t+1}(\boldsymbol{\omega}_{r,t+1}) = \widetilde{T}_{r,t}(\boldsymbol{\omega}_{r,t})  + \frac{1}{C}\Bigg(\widetilde{q}^{p2hh}_{r,t}(\boldsymbol{\omega}_{r,t}) - \widetilde{q}^{exc}_{r,t}(\boldsymbol{\omega}_{r,t})  \\ - \frac{1}{R^{eqv}}(\widetilde{T}_{r,t}(\boldsymbol{\omega}_{r,t}) - T_a)\Bigg), \forall r, 2\leq t \leq |t|-1  
\end{split}
\end{equation}
\begin{equation} \label{eq:temp25min}
\begin{split}
       \underset{D \in \mathcal{P}_{rt}}{\min} \mathbb{P}[  \widetilde{T}_{r,t}(\boldsymbol{\omega}_{r,t}) + \frac{1}{C}\Bigg(\widetilde{q}^{p2hh}_{r,t}(\boldsymbol{\omega}_{r,t}) - \widetilde{q}^{exc}_{r,t}(\boldsymbol{\omega}_{r,t})  - \frac{1}{R^{eqv}}\\(\widetilde{T}_{r,t}(\boldsymbol{\omega}_{r,t}) - T_a)\Bigg)  \geq T_{min}]  \geq 1-\epsilon, \forall r, t = |t|
\end{split}
\end{equation}
\begin{equation} \label{eq:temp25max}
\begin{split}
   \underset{D \in \mathcal{P}_{rt}}{\min} \mathbb{P}[  \widetilde{T}_{r,t}(\boldsymbol{\omega}_{r,t}) + \frac{1}{C}\Bigg(\widetilde{q}^{p2hh}_{r,t}(\boldsymbol{\omega}_{r,t}) - \widetilde{q}^{exc}_{r,t}(\boldsymbol{\omega}_{r,t})  - \frac{1}{R^{eqv}}\\(\widetilde{T}_{r,t}(\boldsymbol{\omega}_{r,t}) - T_a)\Bigg) \leq T_{max}]  \geq 1-\epsilon, \forall r, t = |t| 
   \end{split}
\end{equation}
\end{subequations}

Minimum and maximum temperature limits of the P2HH are enforced in (\ref{eq:tempRangeLow}) and (\ref{eq:tempRangeHigh}) respectively as distributionally robust chance constraints. Temperature evolution of the P2HH electrolyte is imposed by (\ref{eq:tempEvolution0}) and (\ref{eq:tempEvolution}), where $C$ is the specific heat capacity of the electrolysis cell. The last term 
$\frac{1}{R^{e q v}}\left(T_{r, t}-T_{a}\right)$ refers to heat dissipation of the electrolyte, where $R^{e q v} $ (\textdegree{C}/MW) is the equivalent thermal resistance of the electrolyte. Distributionally robust chance constraints (\ref{eq:temp25min}) and (\ref{eq:temp25max}) enforce the temperature limits at the end of representative days.

\begin{subequations}\label{eq:tank}
\begin{equation}\label{eq:tankt1}
    \widetilde{m}^{H2}_{r,t}(\boldsymbol{\omega}_{r,t}) =  \widetilde{n}^{H2}_{r,t}(\boldsymbol{\omega}_{r,t}), \forall r, t = 1
\end{equation}
\begin{equation}\label{eq:tankt2}
    \widetilde{m}^{H2}_{r,t}(\boldsymbol{\omega}_{r,t}) = \widetilde{m}^{H2}_{r,t-1} (\boldsymbol{\omega}_{r,t-1}) +\widetilde{n}^{H2}_{r,t}(\boldsymbol{\omega}_{r,t}), \forall r, t \geq 2
\end{equation}
\begin{equation}\label{eq:tankend}
    \underset{D \in \mathcal{P}_{rt}}{\min} \mathbb{P}[ 
    \widetilde{m}^{H2}_{r,t} (\boldsymbol{\omega}_{r,t}) \leq m^{tank} ] \geq 1-\epsilon, \forall r, t = |t| 
\end{equation}
\end{subequations}

Constraint (\ref{eq:tank}) are associated with the hydrogen tank operation. (\ref{eq:tankt1})-(\ref{eq:tankt2}) impose the hydrogen content evolution in the tank, while (\ref{eq:tankend}) limits the hydrogen content within the tank capacity at the end of each representative days. (\ref{converter})-(\ref{comp}) pertain to capacity limits of the AC/DC converter and compressor respectively.

\begin{subequations}\label{eq:convcomp}
\begin{equation}\label{converter}
    \underset{D \in \mathcal{P}_{rt}}{\min} \mathbb{P}[ 
    \widetilde{p}^{P2HH}_{r,t}(\boldsymbol{\omega}_{r,t}) \leq P^{conv} \eta^{conv} ] \geq 1-\epsilon, \forall r, \forall t  
\end{equation}
\begin{equation}\label{comp}
    \underset{D \in \mathcal{P}_{rt}}{\min} \mathbb{P}[ 
    \widetilde{n}^{H2}_{r,t}(\boldsymbol{\omega}_{r,t}) \leq m^{comp} ] \geq 1-\epsilon, \forall r, \forall t  
\end{equation}
\end{subequations}

\subsection{Linear Decision Rule}
The infinite-dimensional nature of the problem (\ref{eq:obj})-(\ref{eq:convcomp}) as the recourse variables are functions of uncertain parameters that are only revealed in the real time results in an intractable optimization problem. To enable solvability of the problem, linear decision rules \cite{Pourahmadi2019, Ratha2020} are applied, where recourse actions of the flexibility sources (i.e., CHP operations, P2HH operations, electric boiler operations, grid transmission) are approximated as affine responses to the uncertainty realization. The proposed linear decision rules, although somewhat limiting by not covering the dynamic nature of power system operations, provide a straightforward understanding of uncertainty handling in power systems and a reasonable approximation of recourse actions and most importantly, a tractable reformulation of the proposed  problem (\ref{eq:obj})-(\ref{eq:convcomp}) at a lower level of complexity.

The linear recourse actions of the flexibility sources are presented in \ref{app:LDR}. In addition to their nominal scheduling, the flexible agents are assigned optimal affine resposes, which govern their operations in response to uncertainty realization in the real time. Using transmission power as an example, $\widetilde{p}_{r,t}^{trans}(\boldsymbol{\omega}_{r,t})$ refers to the transmitted power in the real-time operation as a function of wind forecast errors $\boldsymbol{\omega}_{r,t}$, $p_{r,t}^{trans}$ refers to the nominal scheduling in the absence of forecast errors and $\beta_{r,t}(\mathbf{1}^{\top} \boldsymbol{\omega}_{r,t})$ denotes the affine response, where $\mathbf{1}^{\top} \boldsymbol{\omega}_{r,t}$ denotes aggregated forecast error and $\beta_{r,t}$ is the affine response parameter (also called participation factor).

\subsection{Model Reformulation}
\begin{figure*}[tbp]
    \centering
    \includegraphics[width = 0.8\linewidth]{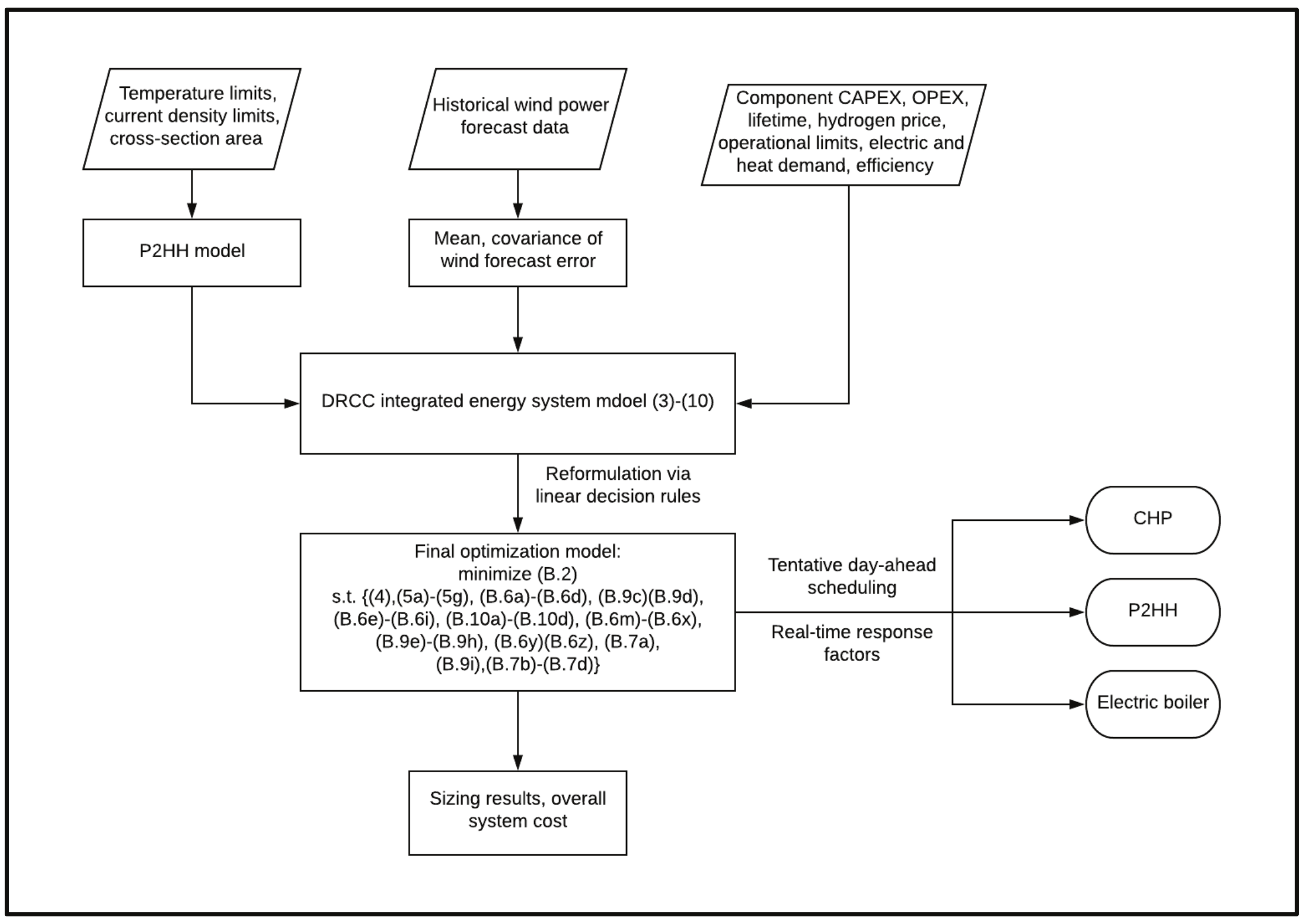}
    \caption{Overall methodology, code and data open source at \cite{SenDRCC, SenTHH}}
    \label{fig:method}
\end{figure*}

Using above linear decision rules, the problem (\ref{eq:obj})-(\ref{eq:convcomp}) can be reformulated as a tractable optimization problem, where the objective (\ref{eq:obj}) is reformulated as a single-level minimization problem and the distributionally robust chance constraints are reformulated as second-order cone constraints. The details are presented as in \ref{app:reformulation}. The overall methodology of this study is shown in Figure \ref{fig:method}.

\subsection{Benchmark Chance-Constrained Model}
A benchmark model using chance-constrained programming (not distributionally robust) is formulated to compare with the developed model. It assumes  the forecast errors follow Gaussian distribution. Similar to distributionally robust chance constraints, chance constraints assuming Gaussian distribution can also be reformulated into second-order cone constraints. The derivation is seen in \ref{app:gaussian}.

\subsection{Assumptions}
Throughout this study, the following assumptions are made.
\begin{itemize}
    \item The probability distribution of wind power uncertainties is assumed unknown, which motivates the development of a distributionally robust chance-constrained model. 
    \item The exact mean and covariance of wind forecast errors are assumed estimated from historical data.
    \item The recourse actions of flexibility resources in the real-time are assumed linearly related to wind power uncertainty realization.
    \item The ON/OFF status of the CHP plant is assumed determined in the day-ahead, while its production levels can be adjusted in the real time according to wind power production.
    \item Hydrogen produced from alkaline electrolysers is assumed fully sold at constant price.
    \item The network constraints are not yet included.
\end{itemize}

%% file: 4.Case.tex
\section{Case Description}\label{sec:case}
A case study is performed on an integrated energy system to test the effectiveness of the developed distributionally robust chance-constrained model for the power to heat facility sizing and IES operations. The integrated energy system includes electric and heat loads. Some technical details of the case study are introduced below, while the others can be accessed on the online open-source repository \cite{SenDRCC}. The models are formulated with YALMIP toolbox \cite{Lofberg2004} on Matlab and solved with the Gurobi solver 9.0 \cite{gurobi}. The optimization models take 2-3 hours to run on an Intel Core i5-9300H CPU running at 2.4GHz with an 8GB RAM.

\subsection{CHP Plant}
\begin{figure}[h]
    \centering
    \includegraphics[width = 0.9\linewidth]{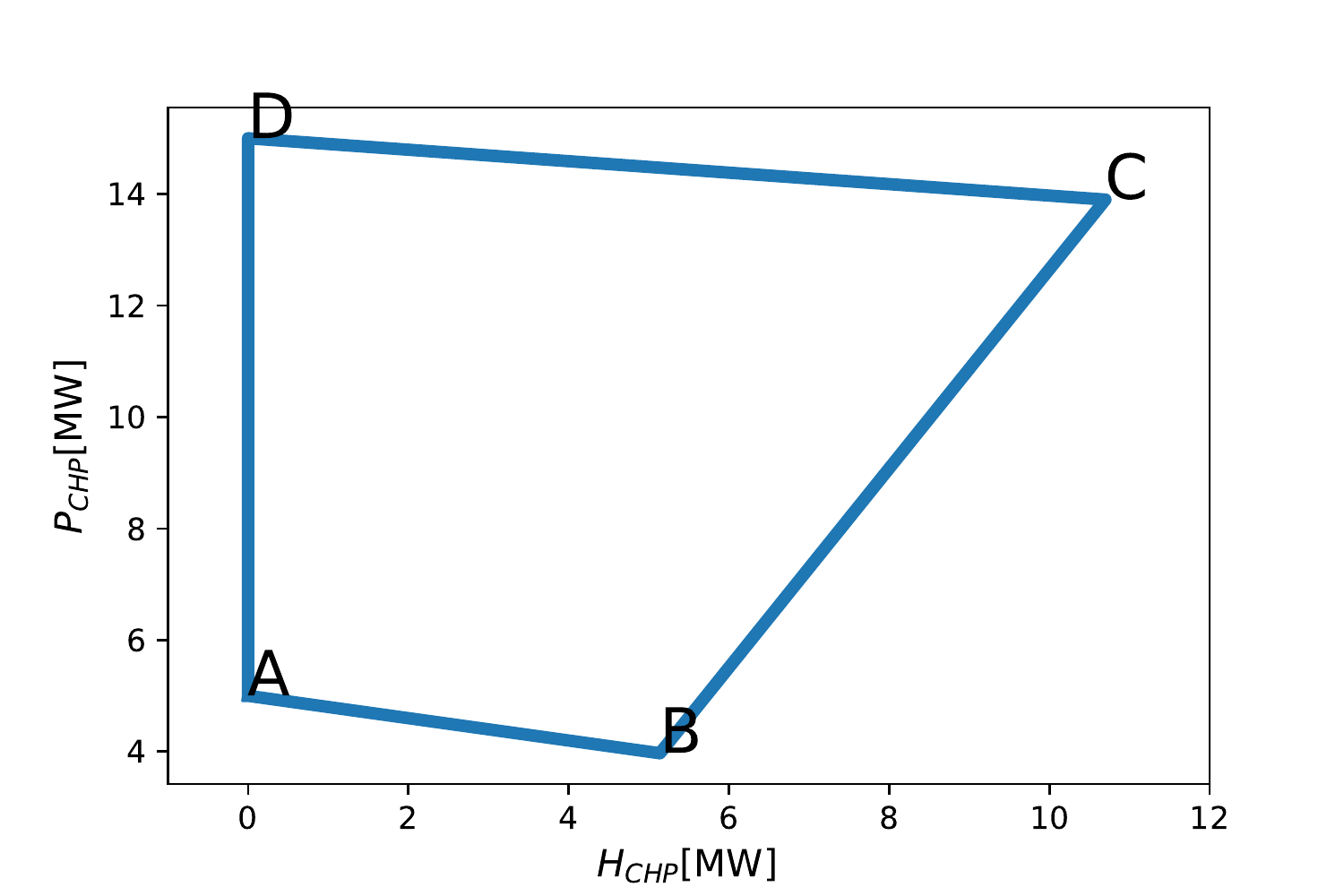}
    \caption{Operational region of the CHP plant}
    \label{fig:CHPoperationalRegion}
\end{figure}

For CHP plants, electricity and heat production are coupling. A convex operation region similar to Figure \ref{fig:CHPoperationalRegion} has been extensively applied in energy system studies \cite{Li2019,Chen2018, Chen2015, Pan2020} to characterize CHP plant operation. In our study,  non-negative variables $\mathbf{\widetilde{x}}$ summing up to 1 are assigned to each corner point to represent weights given to each corner point during CHP operation.

\subsection{Electric and Heat Demands}
\begin{figure}[h]
    \centering
    \includegraphics[width=\linewidth]{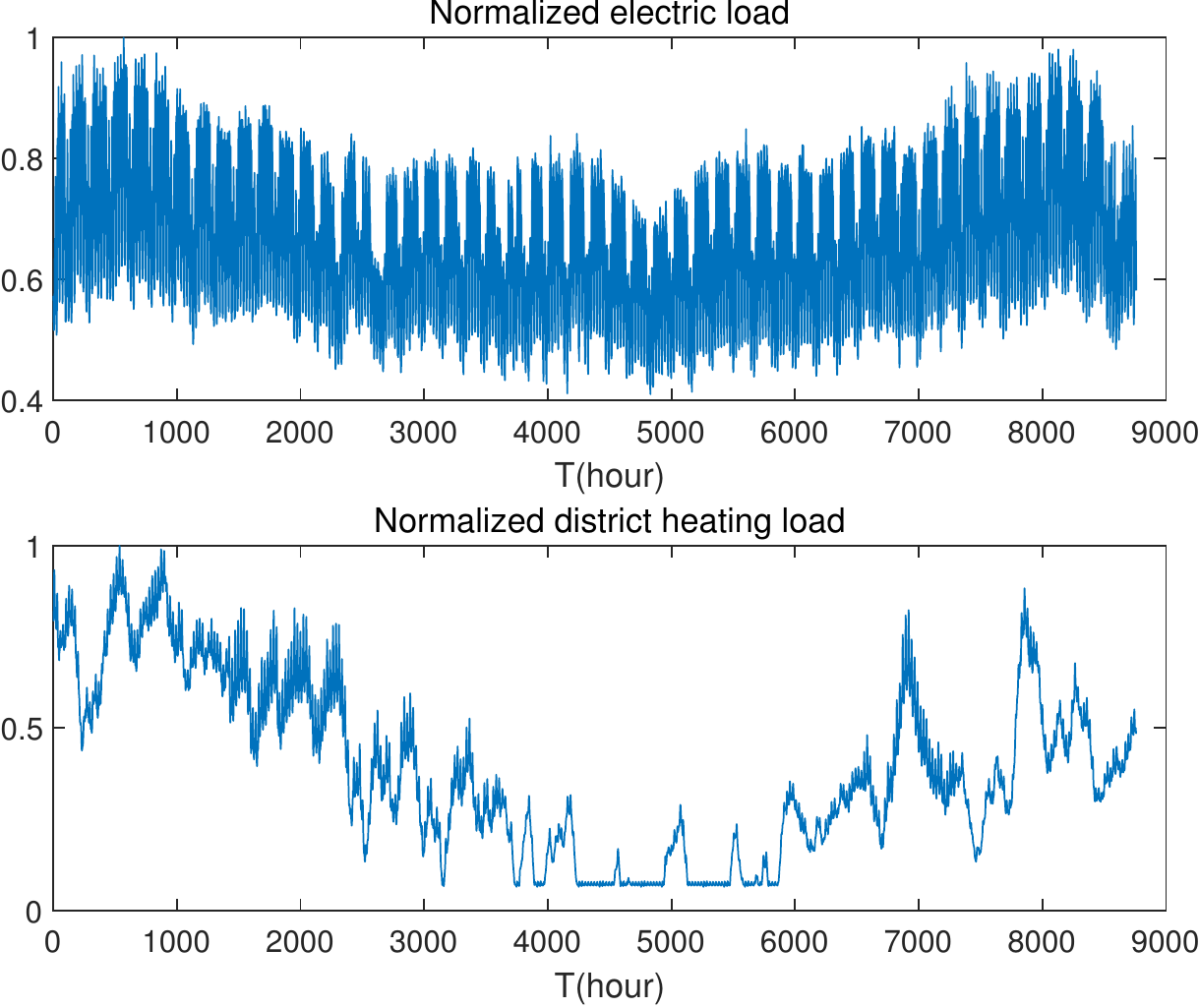}
    \caption{Normalized hourly electric and district heating demands}
    \label{fig:hourlyLoads}
\end{figure}

Yearly electric and heat demands are normalized from \cite{nordpool} and \cite{kunz_friedrich_2017_1044463} respectively, shown as Figure \ref{fig:hourlyLoads}. While electric load and district heating load both present seasonal variations, district heating load
variation is more fluctuating. Much less heat is consumed during the summer than the winter. 

In order to capture the load characteristics while maintaining computational tractability, a K-means clustering algorithm is applied to form a set of representative days for the target year. This technique has found its wide application in energy system studies, e.g., \cite{Pourahmadi2019} to cluster various types of scenarios. The underlying idea for K-means clustering technique is to cluster scenarios that are close in terms of Euclidean distance in high-dimensional space. In this study, we limit us to 10 representative days for the target year since further increasing the number of scenarios does not significantly change the results. 

\subsection{Wind Uncertainties}
\begin{figure}[h]
    \centering
    \includegraphics[width=\linewidth]{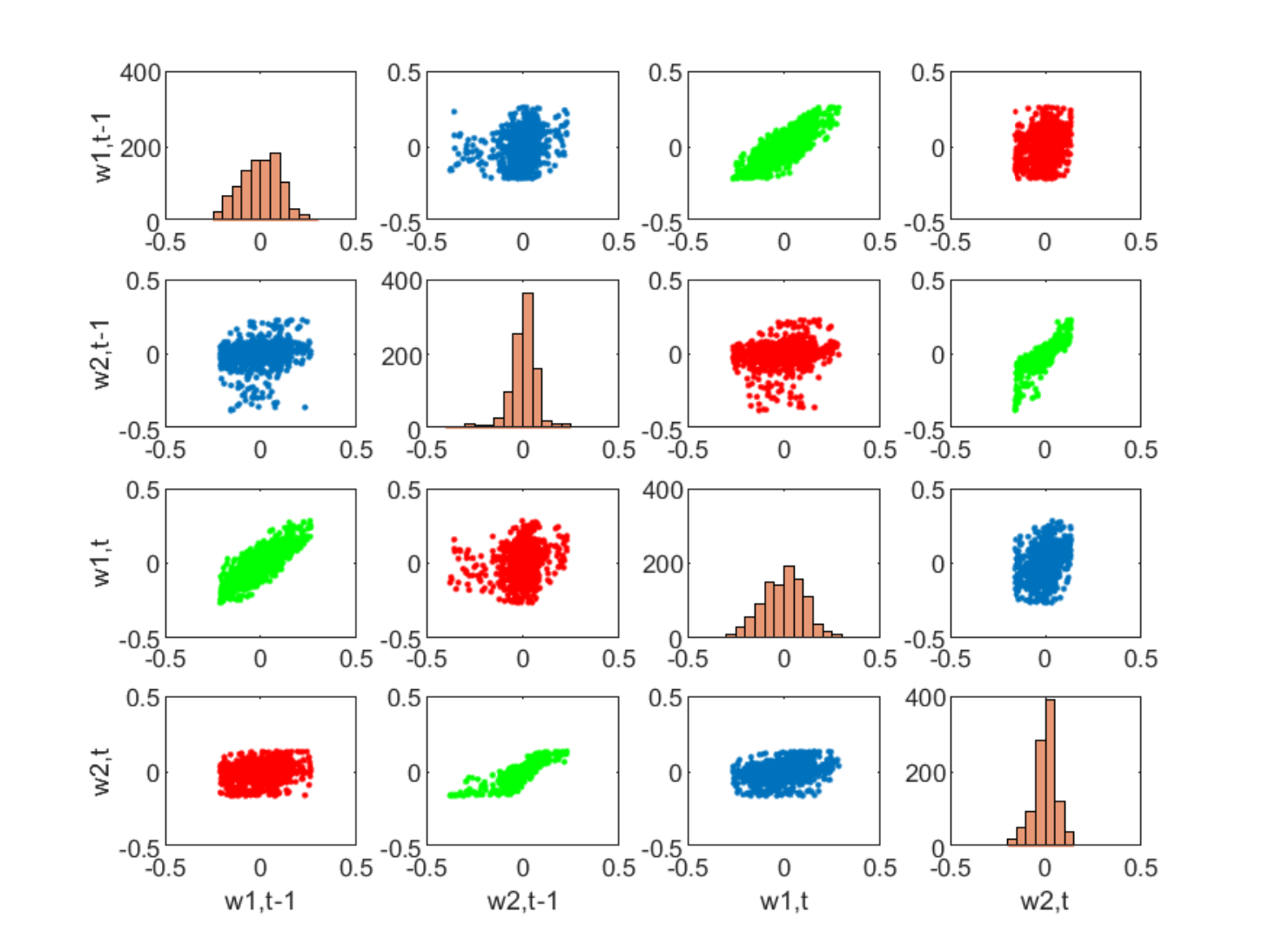}
    \caption{Wind forecast errors of wind farm 1 (w1) and wind farm 2 (w2) at two consecutive hours (t-1,t) in per-unit}
    \label{fig:DRCC}
\end{figure}

\begin{figure}[h]
    \centering
    \includegraphics[width=\linewidth]{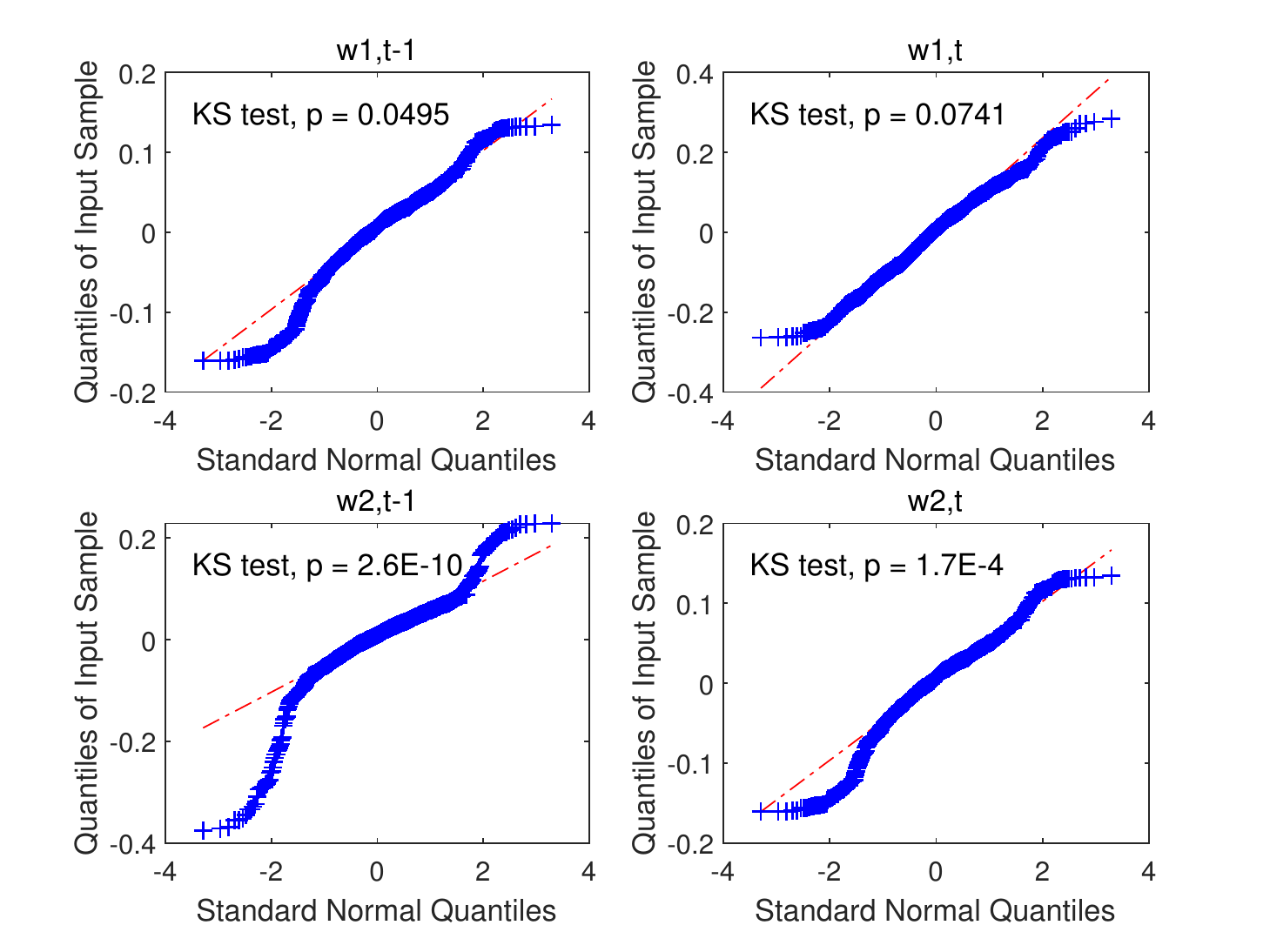}
    \caption{Quantile-quantile plots and p values for wind forecast errors of wind farm 1 (w1) and wind farm 2 (w2) at two consecutive hours (t-1,t)}
    \label{fig:DRCC1}
\end{figure}

Uncertainty handling has been an important consideration in energy system models. In this study, developed a distributionally robust chance-constrained model to account for wind uncertainties. Linear decision rules are applied to represent real-time operation when wind uncertainty is revealed. This provides an intuitive and straightforward understanding of uncertainty handling in power systems. 

In order to implement this distributionally robust chance-constrained planning model, mean of wind forecast as well as covariance of wind forecast error vector for each representative day and hour are required. For this purpose, 1000 wind scenarios each containing 10 representative days' wind forecast profile are directly acquired from the dataset of \cite{Pourahmadi2019} to calculate these parameters, i.e., $\mathbf{m}_{r,t}$, $\boldsymbol{\Sigma}_{r,t}$. To better illustrate the intention of using DRCC planning, Figure \ref{fig:DRCC} presents wind forecast errors for two wind farms at two consecutive hours in a representative day. Diagonal histograms show frequency distributions of wind forecast errors, which have a mean of 0. Off-diagonal plots illustrate the spatial and temporal wind forecast error correlation between two wind farms for two consecutive hours. Specifically, plots in blue show spatial correlation of the two wind farms at the same hour, plots in green show temporal correlation for the same wind farm, while plots in red show both temporal and spatial correlation of wind forecast errors. It can be observed from these plots that these forecast error correlations do not necessarily match any specific type of distribution. In addition, normality tests are performed based on the quantile-quantile plot and the Kolmogorov–Smirnov (KS) test, shown in Figure \ref{fig:DRCC1}. On all plots systematic departure from the straight line is observed on the tails. P values from the KS test under the null hypothesis that the wind forecast errors follow the normal distribution are also shown. At the 10\% significance level, the null hypothesis is rejected for all, which indicates the forecast errors do not fit into normal distribution. These point out the significance of using distributionally robust optimization models to address wind forecast error uncertainties.

%% file: 5.Results.tex
\section{Results}\label{sec:results}
\begin{figure}[h]
    \centering
    \includegraphics[width = \linewidth]{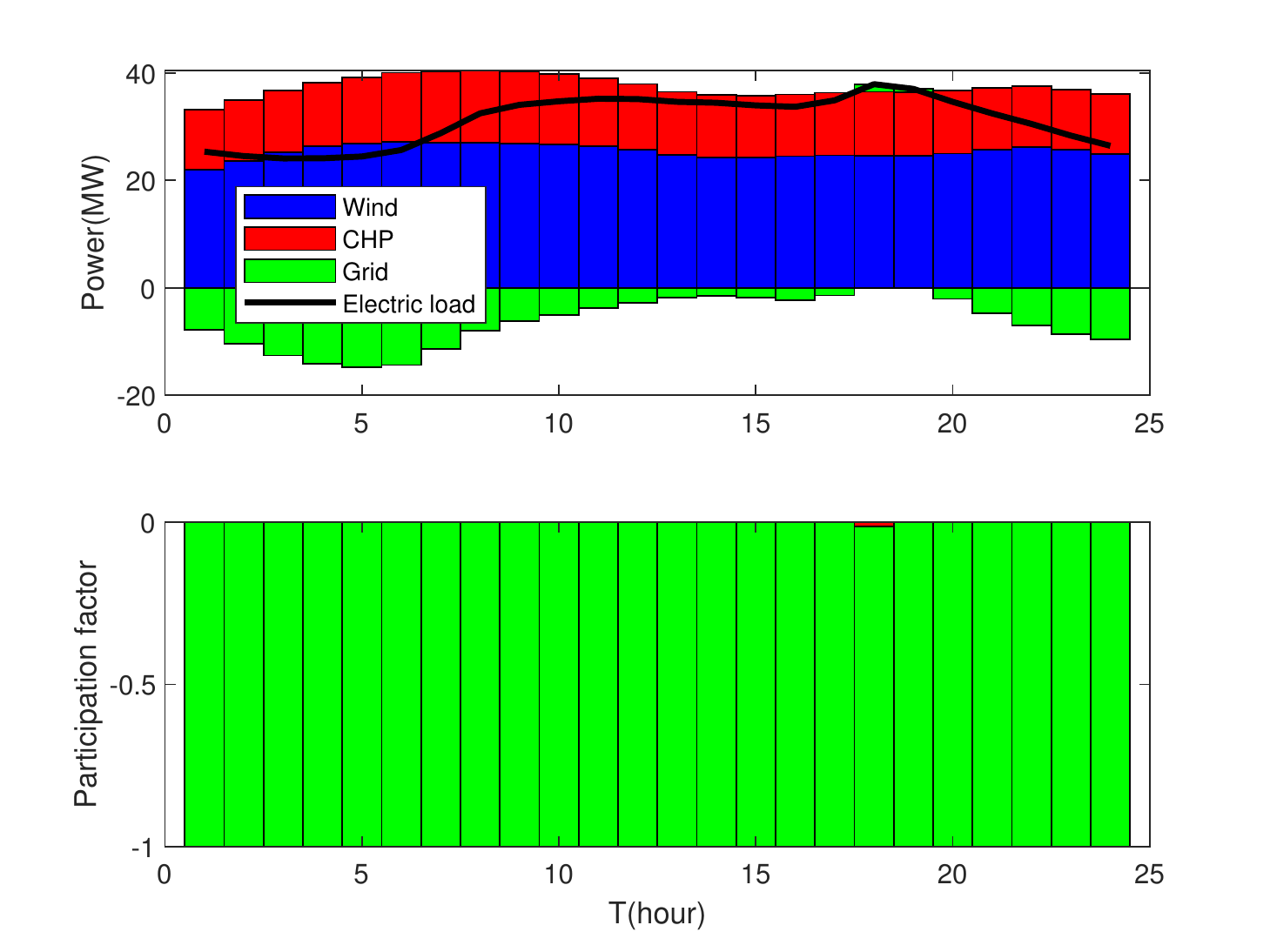}
    \caption{Upper: nominal scheduling of wind turbines, CHP plant and transmitted power. Lower: Stochastic scheduling of transmitted power and CHP plant responding to wind power uncertainties, $\epsilon$ = 0.05}
    \label{fig:CHPpowerbalance}
\end{figure}

\begin{figure}[h]
    \centering
    \includegraphics[width = .9\linewidth]{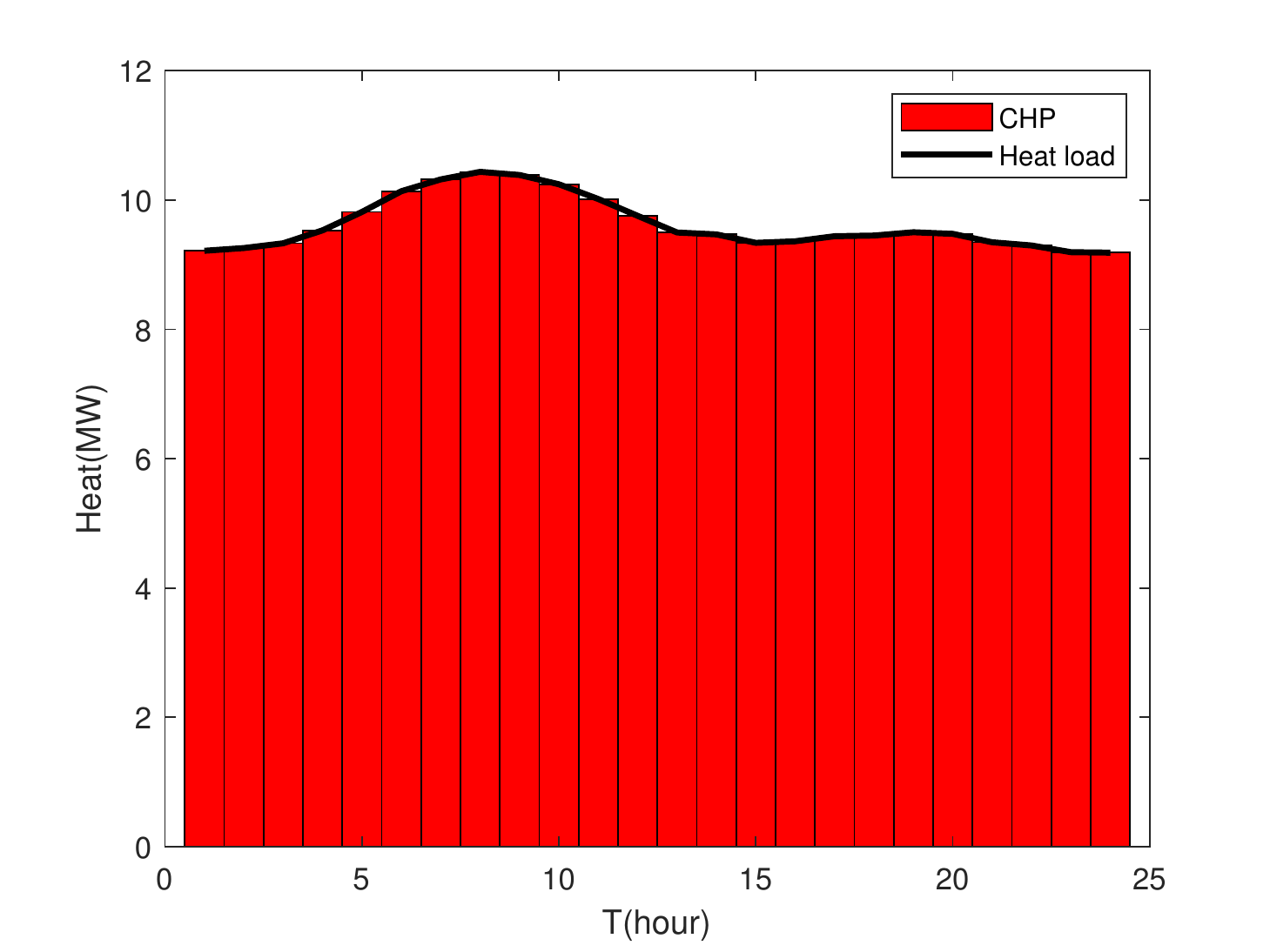}
    \caption{Nominal scheduling of CHP plant, $\epsilon$ = 0.05}
    \label{fig:CHPheatbalance}
\end{figure}

This section covers important findings from the above case study. Specifically, section \ref{effec} verifies the effectiveness of the developed model by looking at power and heat balance of the integrated energy system, as well as P2HH operation. Section \ref{P2H} discusses the effects of introducing P2HH and electric boiler to the IES in terms of economical performance, flexible CHP plant operation and reduced inverse power flow. Section \ref{confi} deals with the system performance at various confidence levels. Section \ref{cc} compares the distributionally robust chance-constrained model with a chance-constrained model assuming Gaussian distribution. Section \ref{rev} looks at the profit distribution across various stakeholders, i.e., wind plants, CHP plants and P2HH/EB investors. 

\subsection{System Operations}\label{effec}
\subsubsection{Power and Heat Balance}

\begin{figure}[h]
    \centering
    \includegraphics[width = \linewidth]{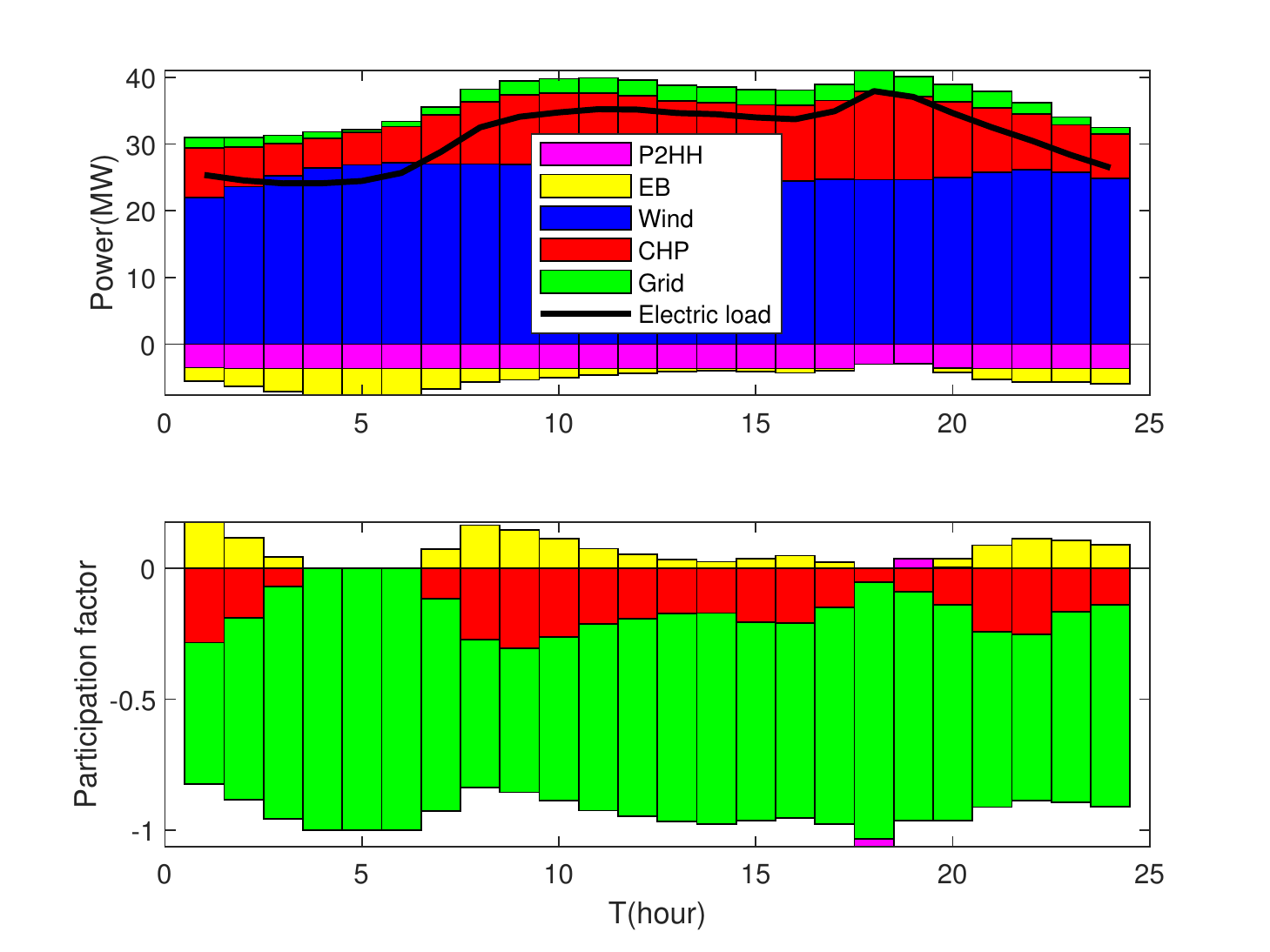}
    \caption{Nominal and stochastic scheduling of various electricity generators/users, $\epsilon$ = 0.05}
    \label{fig:P2HHEBelecbalance}
\end{figure}

\begin{figure}[h]
    \centering
    \includegraphics[width = \linewidth]{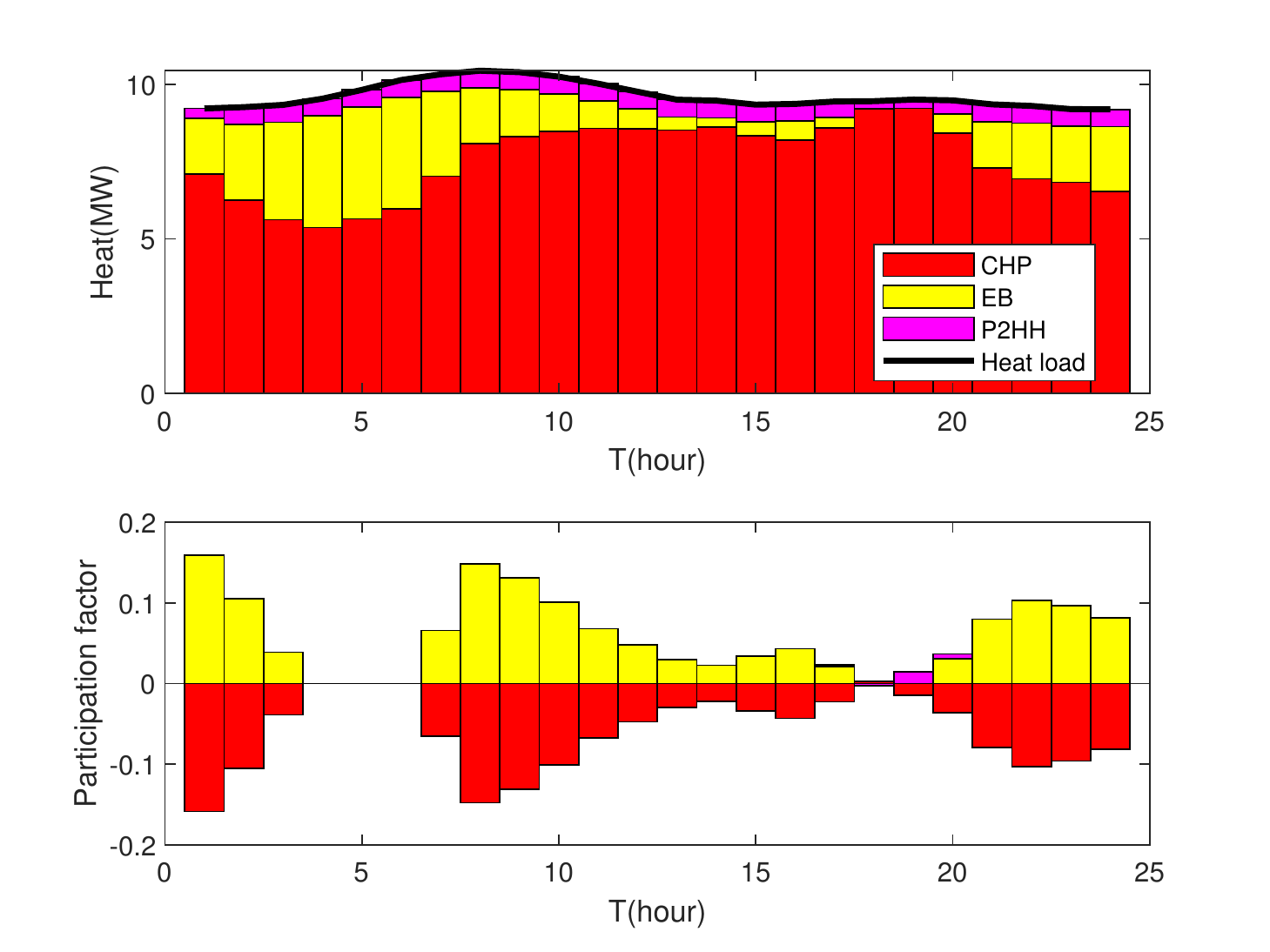}
    \caption{Nominal and stochastic scheduling of various heat generators, $\epsilon$ = 0.05}
    \label{fig:P2HHEBheatbalance}
\end{figure}

Figure \ref{fig:CHPpowerbalance} presents the nominal scheduling and affine policy parameters (participation factors) of wind generators, CHP plant and transmission grid under the existing integrated energy system context for a representative day with high heating demand, where the P2HH facility and electric boiler are not available. The inelastic electric load is met mostly from wind power. Even in some hours (e.g., hour 4-6), wind power exceeds the electric loads. However, the CHP plant is held online throughout the day, producing a nearly constant amount of power (around 15 MW), which results from the binding high heating demand. A large amount of power is
thus transmitted into the high-voltage transmission network for local balancing, resulting in inverse power flow challenging grid operation. The lower plot in Figure \ref{fig:CHPpowerbalance} indicates that the transmission grid is the only source of flexibility for the distribution network to respond to the uncertain wind power at most of the hours due to the fact that the CHP plant lacks operational flexibility when facing high heating load. Similarly, Figure \ref{fig:CHPheatbalance} shows the nominal heating scheduling of the CHP plant, which is able to meet the high heating demand. However, as it is the only heating source, it is not able to adjust its heat production when facing wind forecast errors.

On the other side, Figure \ref{fig:P2HHEBelecbalance} and Figure \ref{fig:P2HHEBheatbalance} present the nominal scheduling and affine responses of various agents in the integrated energy system for power and heat respectively where the P2HH facility and electric boiler are introduced. It is observed in Figure \ref{fig:P2HHEBelecbalance} that the inverse power flow is eliminated, while a small amount of power is transmitted into the system to meet extra demand from power to heat facilities. The CHP plant regains operational flexibility as seen in Figure \ref{fig:P2HHEBelecbalance} where it is able to lower its electricity production when unforeseen extra wind power is injected in the system.
The electric boiler also responds to wind uncertainties by adjusting its power consumption, e.g., increasing consumption when facing extra wind power. It is worth mentioning that the P2HH facility is working steadily at around its full capacity in order to gain more profits by selling hydrogen, thus far less flexible than the electric boiler. It is also noticed that the transmission grid still takes an important role in local balancing due to power to heat facilities' relatively small capacities. On the heating side, the system gains two extra heating sources: the P2HH facility and electric boiler to satisfy the heating demand. The electric boiler responds to unforeseen wind power by consuming more energy, and the CHP plant decreases its heat production.

\subsubsection{P2HH Operations}
\begin{figure}[t]
    \centering
    \includegraphics[width = \linewidth]{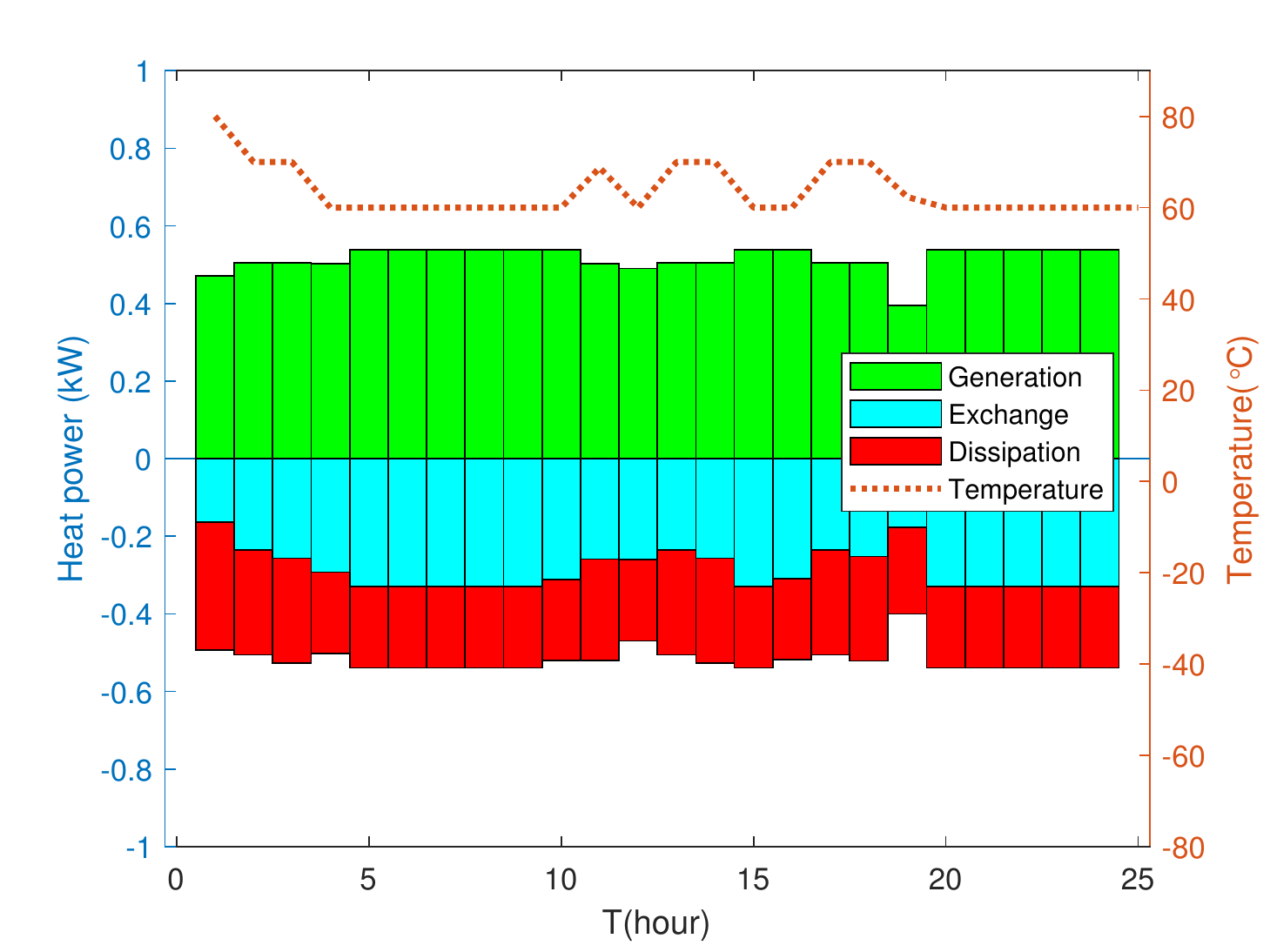}
    \caption{Temperature evolution for a representative day}
    \label{fig:tempEvol}
\end{figure}

Figure \ref{fig:tempEvol} validates the P2HH model by looking at temperature evolution for a typical day. The cell temperature evolves according to its heat generation and release. When the heat generation exceeds the sum of district heating use and dissipation, the cell temperature increases. When the heat generation is lower than that, the cell temperature drops. However, the alkaline electrolysis cell temperature only evolves in a small range of 60-80\textdegree{C} referring to \cite{Dincer2014}. It is also noticeable that heat dissipation accounts for a high percentage of heat release due to that the cell is operating at a relatively high temperature compared to the ambient one. In order to further enhance the P2HH facility's overall energy efficiency, proper insulation should be designed.

\subsection{Effects of Introducing P2HH and Electric Boiler}\label{P2H}
In this section, the effects of introducing P2HH infrastructure and electric boiler to the integrated energy system are explored. The analysis is carried out in terms of system cost, CHP plant flexibility and excess power. Four scenarios are considered regarding the involvement of P2HH and eletric boiler, which are listed in Table \ref{tab:scenario}.
\begin{table}[h]
    \centering
        \caption{Simulation scenarios}
    \begin{tabular}{lc}
\hline
Scenarios & \hspace{2cm} System components \\\hline
         Scenario 1& \hspace{2cm} CHP  \\
         Scenario 2& \hspace{2cm}CHP, EB\\
         Scenario 3 &\hspace{2cm}CHP, P2HH\\
         Scenario 4 & \hspace{2cm}CHP, EB, P2HH\\
         \hline
    \end{tabular}
    \label{tab:scenario}
\end{table}

\subsubsection{System Cost}
\begin{table*}[h]
    \caption{Integrated energy system specifications and annualized cost under various scenarios}
    \centering
    \begin{tabular}{lccccc}
    \hline
       Scenarios  &  EB & P2HH & Tank & Compressor &Annualized IES cost\\ \hline
        Sceanrio 1 & - & - & - & - & \$$3.6\times 10^6$\\
          Sceanrio 2 & 6.7MW & - & -&-&\$$1.8\times 10^6$\\
            Sceanrio 3 & - & 3.7MW & 1500kg&67kg/h&\$$4.4\times 10^5$\\
              Scenario 4 & 4.0MW & 3.5MW &1500kg & 64kg/h &\$$-1.7\times 10^5$\\
        \hline
    \end{tabular}
    \label{tab:IEScost}
\end{table*}
Table \ref{tab:IEScost} covers the optimal sizing and annualized system costs of various facilities for the integrated energy system under the 4 scenarios. By introducing a 6.7 MW electric boiler into the existing integrated energy system, the system cost is halved due to reduced CHP output (hence reduced fuel cost) and reduced excess power which could have brought considerable penalty costs to the system. On the other side, by introducing a 3.7MW P2HH facility, the system cost drops by a magnitude, which comes from the extra revenue from using excess power to producing hydrogen, in addition to reduced CHP cost and reduced excess power. By a combination of a 3.5 MW P2HH facility and a 4MW electric boiler, the system starts to make extra profit on top of meeting local electric and heating demands. However, our detailed economic analysis also found that although the system cost is reduced, the profit is distributed unevenly across various stakeholders. This part will be expanded in section \ref{rev}.

\begin{figure}[t]
\begin{subfigure}{.48\linewidth}
  \centering
  \includegraphics[width=\linewidth]{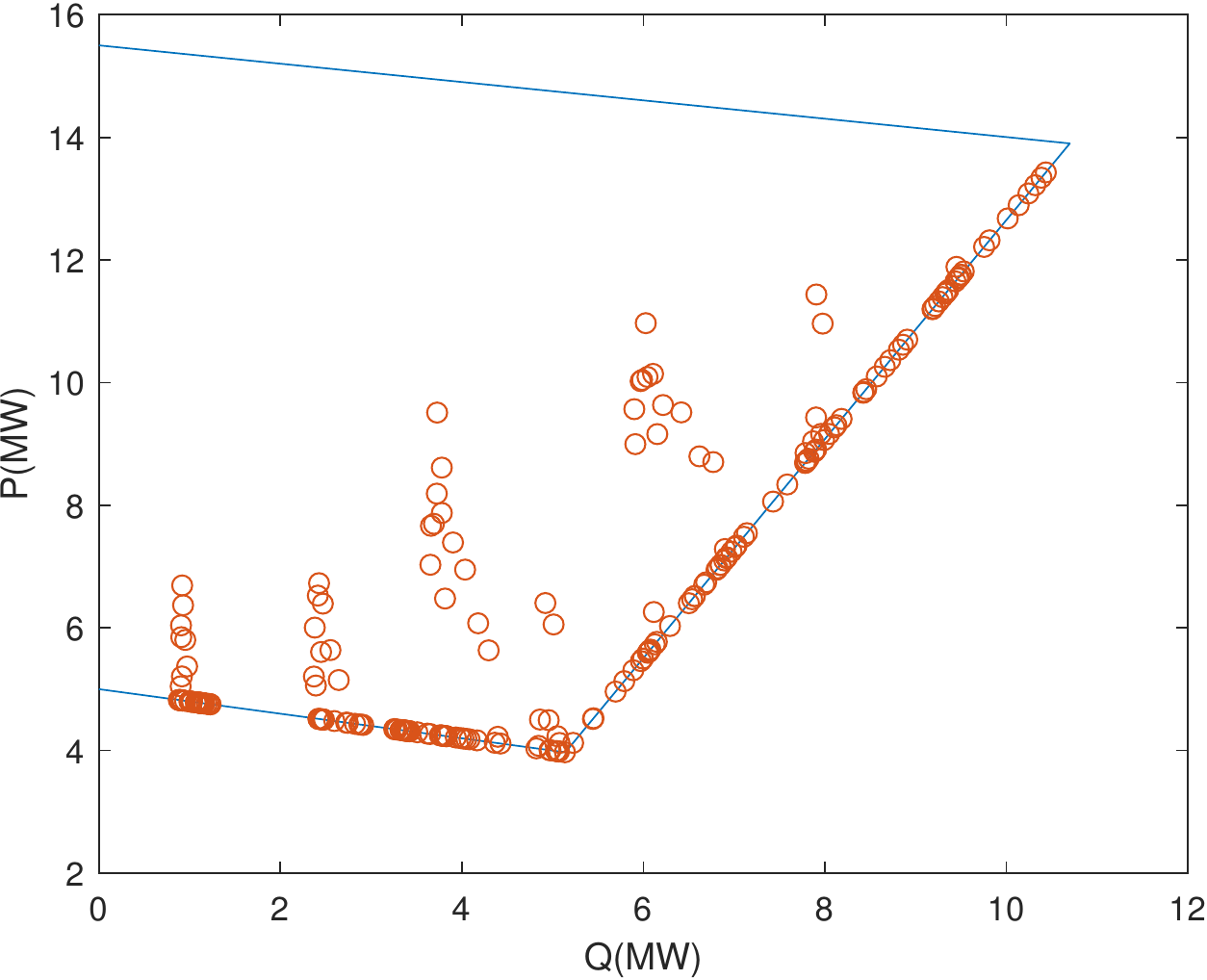} 
  \caption{Scenario 1}
  \label{fig:sce-first}
\end{subfigure}
\begin{subfigure}{.48\linewidth}
  \centering
  \includegraphics[width=\linewidth]{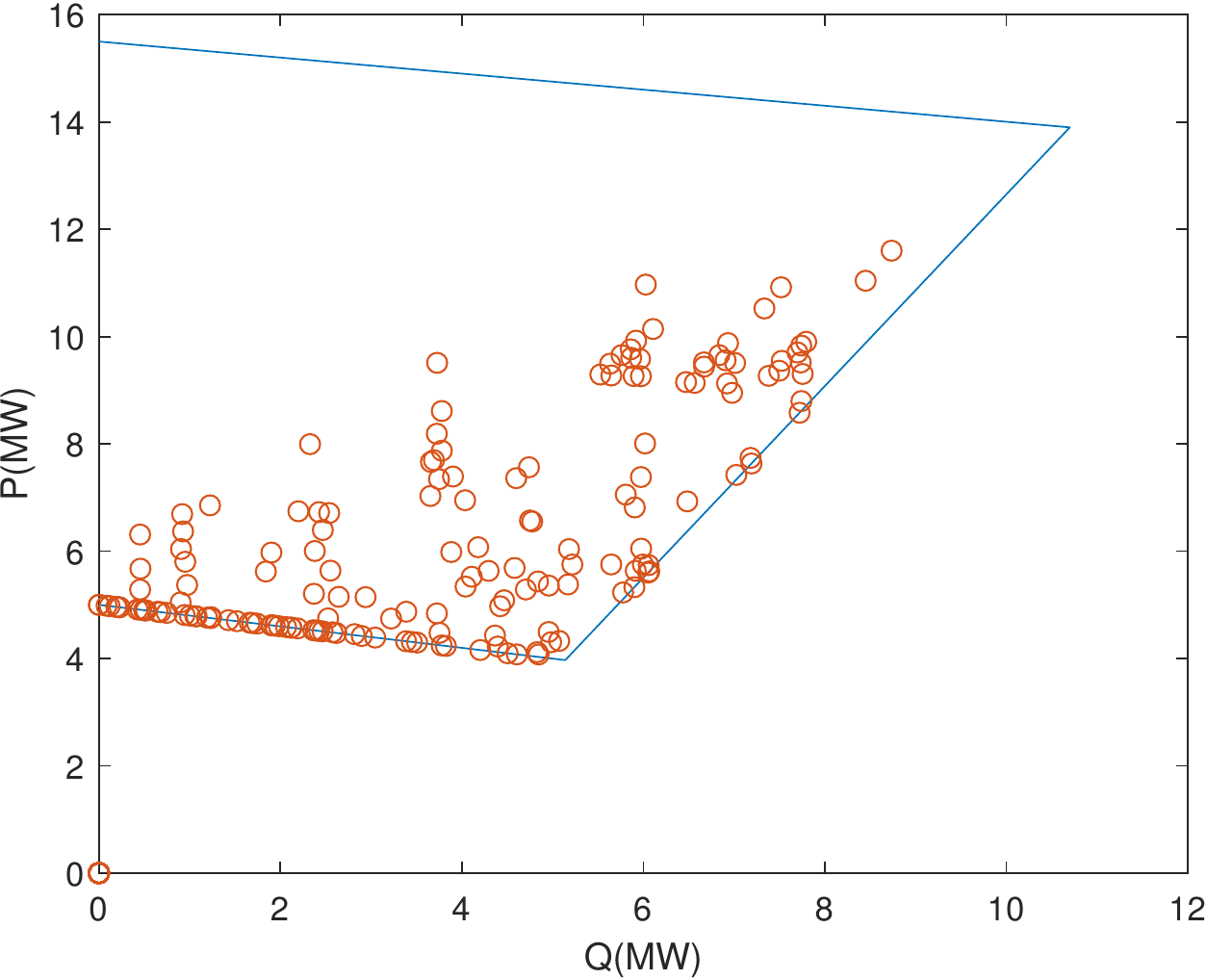} 
  \caption{Scenario 2}
  \label{fig:sce-second}
\end{subfigure}
\newline
\begin{subfigure}{.48\linewidth}
  \centering
  \includegraphics[width=\linewidth]{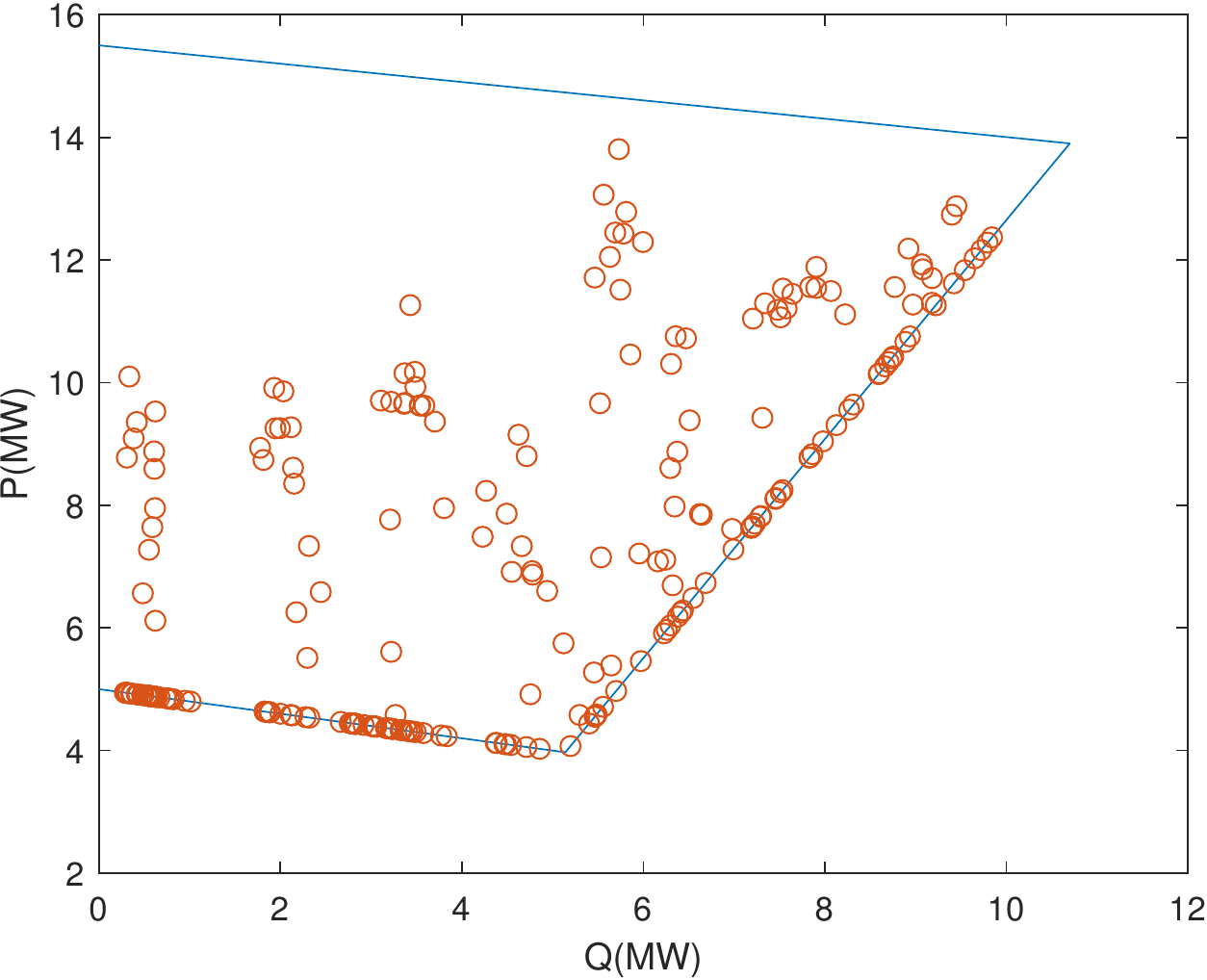} 
  \caption{Scenario 3}
  \label{fig:sce-third}
\end{subfigure}
\begin{subfigure}{.48\linewidth}
  \centering
  \includegraphics[width=\linewidth]{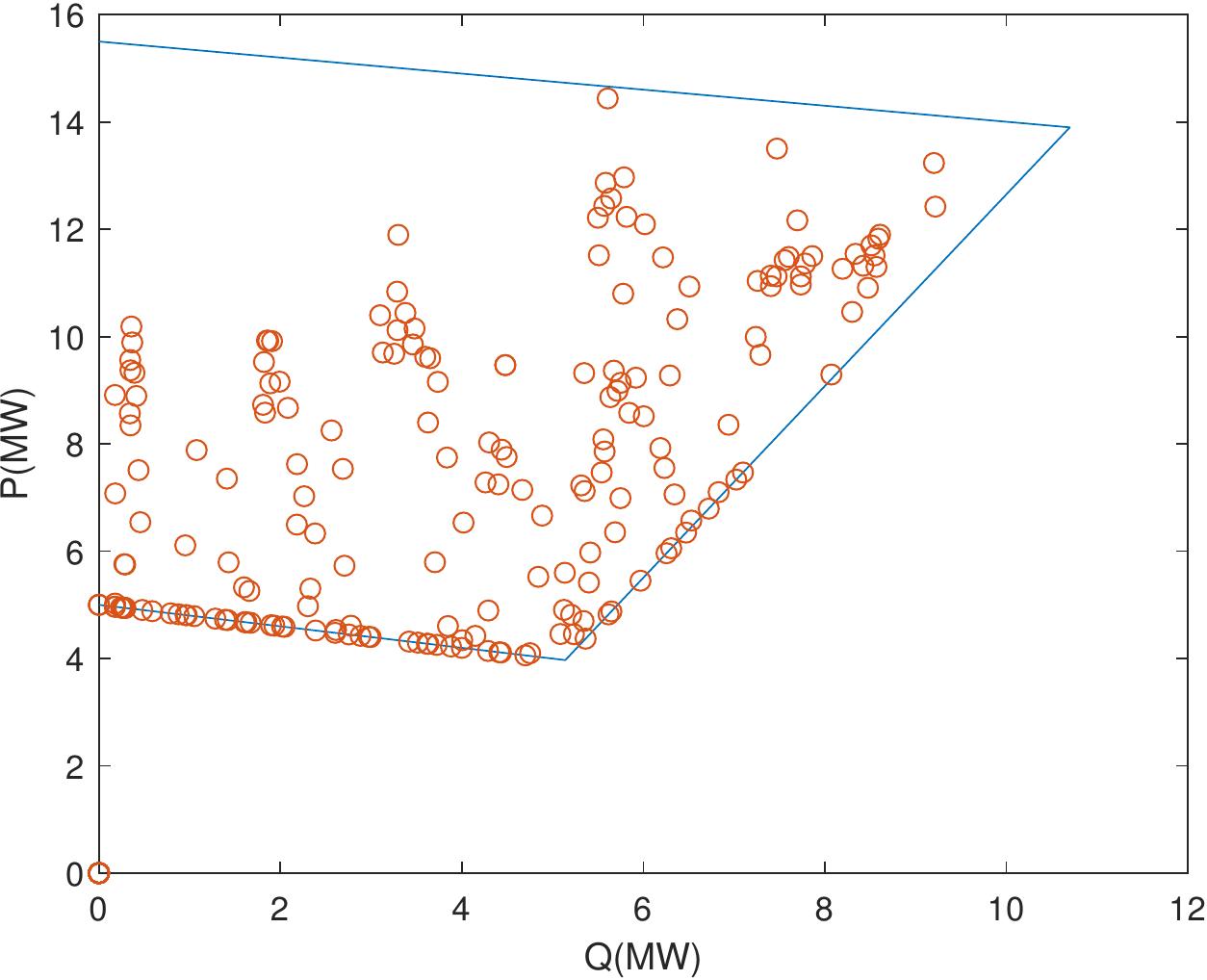} 
  \caption{Scenario 4}
  \label{fig:sce-fourth}
\end{subfigure}
\caption{CHP operations for various scenarios}
\label{fig:CHPoperations}
\end{figure}

\begin{figure}[h]
    \centering
    \includegraphics[width = \linewidth]{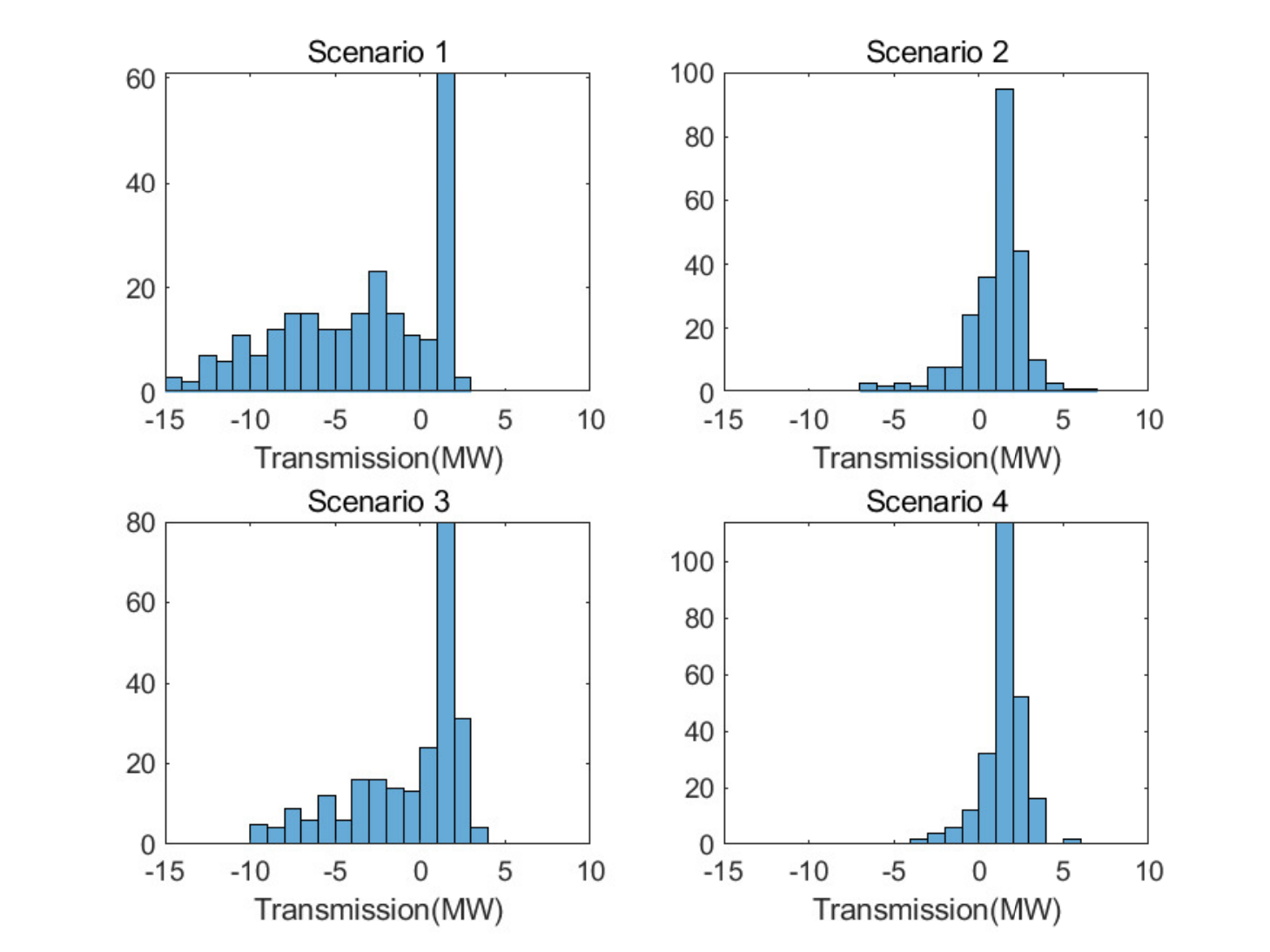}
    \caption{Power transmission under various scenarios}
    \label{fig:trans}
\end{figure}

\subsubsection{Flexible CHP Operations}
The points in Figure \ref{fig:CHPoperations} represent working points of the CHP plant throughout the 10 representative days under the 4 scenarios. For scenario 1 where the CHP plant is the only heat source for the integrated energy system, the CHP plant's working points lie on its right boundary most of the hours, especially when facing high heating demands. Thus the CHP plant lacks operational flexibility. By introducing extra heat sources including P2HH and electric boiler into the system, many of the CHP plant's working points shift out of the right boundary as can be observed in all other scenarios, which indicate that its operational flexibility has been improved. It is also seen that the electric boiler has a better performance compared to the P2HH facility in terms of helping improve CHP plant's operational flexibility, which is attributed to its higher power to heat conversion ratio and larger capacity.

\begin{figure}[t]
    \centering
    \includegraphics[width = \linewidth]{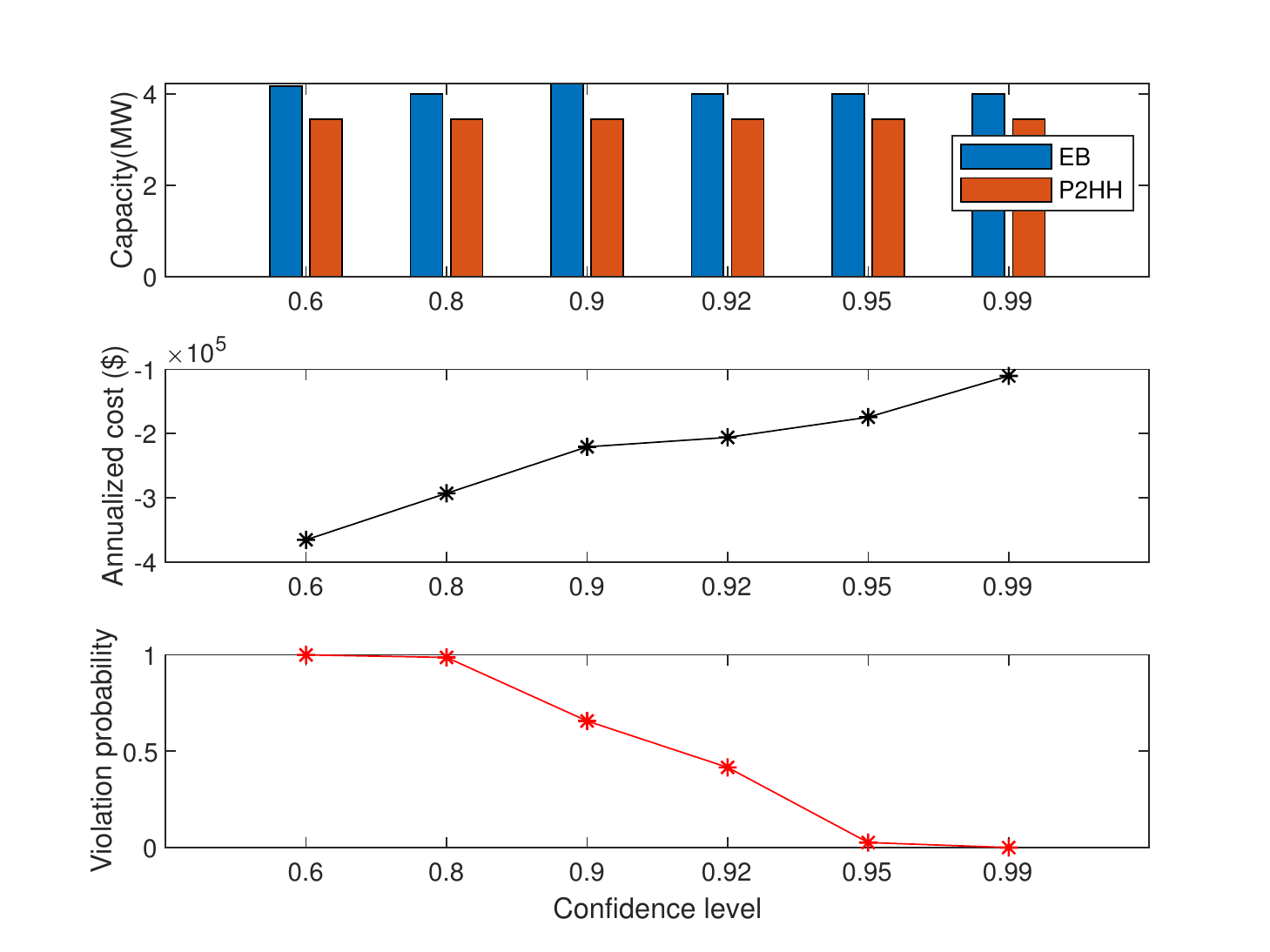}
    \caption{P2HH/EB sizing, system costs and violation probabilities at various confidence levels}
    \label{fig:confidence}
\end{figure}

\subsubsection{Reduced Inverse Power Flow}

\begin{figure}[h]
\centering
    \includegraphics[width = \linewidth]{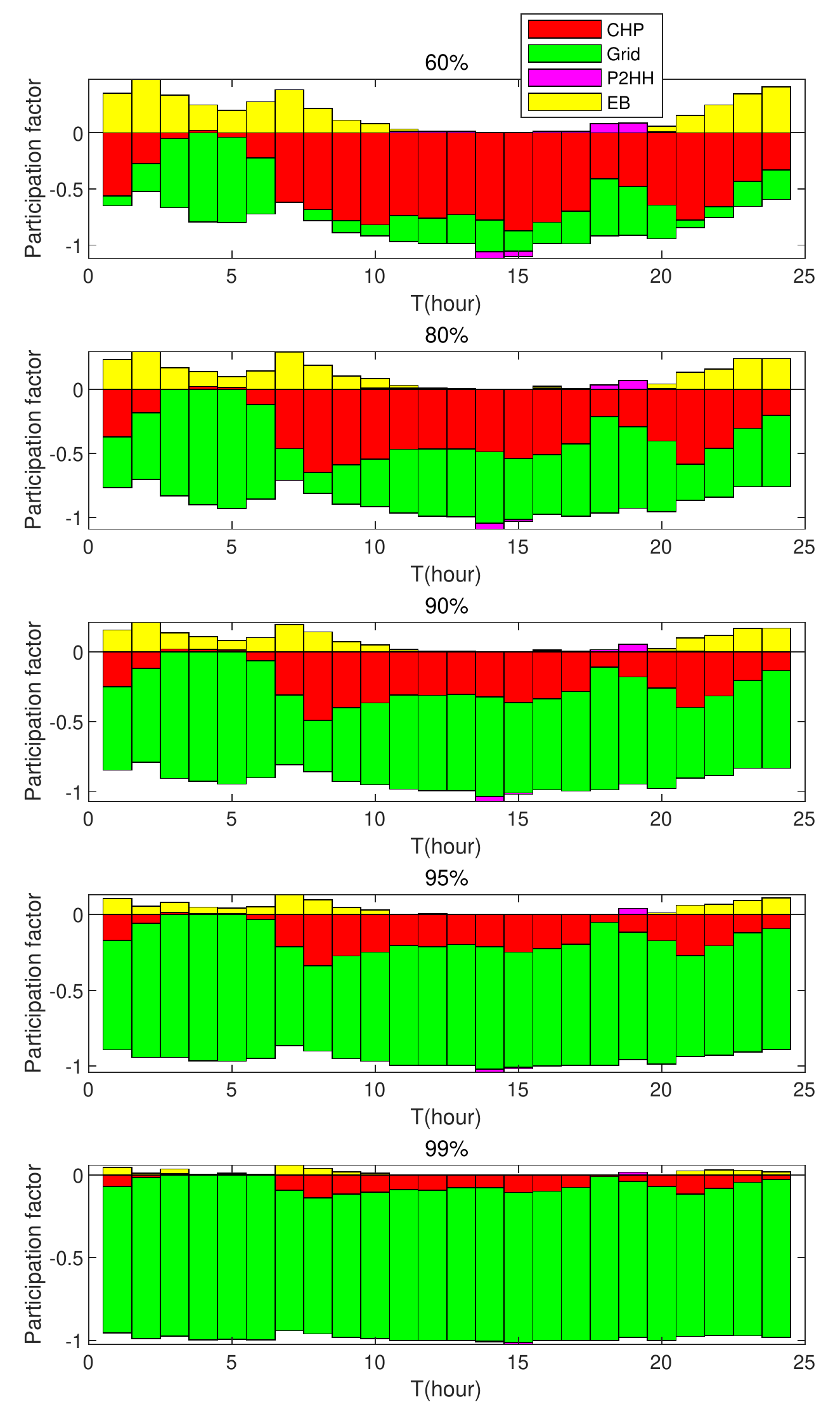}
    \caption{Participation factors at various confidence levels}
    \label{fig:beta}
\end{figure}

Figure \ref{fig:trans} summarizes transmitted power from the high-voltage transmission grid under various scenarios. It is evident that a large amount of inverse power flow is taking place under scenario 1 due to inflexible CHP plant operation. By introducing P2HH and electric boiler, most of the inverse power flow is eliminated. Moreover, the integrated energy system absorbs electricity from the transmission grid to satisfy extra demand from the P2HH facility and electric boiler. It is noticeable that under scenario 2, more inverse power flow is eliminated than scenario 3, which is attributed to the electric boiler's larger capacity and higher power to heat conversion efficiency.

\subsection{Effect of Confidence Levels}\label{confi}
This section looks at the influence of different confidence levels, i.e., $1-\epsilon$ on the system performance. Obviously, a higher confidence level indicates a more stringent model and thus a higher system cost. Figure \ref{fig:confidence} shows that the optimal power to heat facility sizing does not vary significantly under different confidence levels. However, a steady increase in the system cost is observed, which is related to the higher transmission costs at higher confidence levels incurred by expensive regulating services from the transmission grid, shown in Figure \ref{fig:beta} where transmission grid takes up more of the role of handling wind uncertainties with increasing confidence level requirement. The underlying reason is that in order to satisfy the high confidence level requirement, more flexible and reliable resource has to be in position to handle wind uncertainties. The CHP plant, P2HH and electric boiler though can provide certain degree of flexibility, are limited by their capacities. Transmission grid is then in place to cope with wind uncertainties by exchanging electricity with the integrated energy system, which leads to higher transmission costs. 

A higher confidence level implies a higher system cost, but at the same time a more conservative and robust model. In order to verify this, an out-of-sample violation test is performed for the distributionally robust chance-constrained models with different confidence levels. A new dataset is constructed based on classic bootstrapping method. In this test, the affine policies are fixed according to the optimization results for each model with a different confidence level. The 1000 scenarios in the new dataset are fed into the model using the fixed affine policies. The out-of-sample violation probability is defined as the percentage of scenarios where at least one inequality constraint, e.g., (\ref{eq:rampup2}), (\ref{eq:rampdown2}) is violated. The violation probability shows how robustly the wind uncertainty is characterized in the DRO model. A higher violation probability indicates a poor and non-robust characterization. The test results are available in Figure \ref{fig:confidence}, which shows that the violation probability is at a high level when the confidence level is lower than 0.8, which implies that almost under every scenario at least one inequality constraint is violated. When the confidence level exceeds 0.95, the original dataset is represented robustly, with nearly zero violation of the inequality constraints. Taking a balance between model robustness and economical performance, we adopt a confidence level of 0.95 for our application.

\subsection{Comparison with Chance-Constrained Model}\label{cc}
\begin{table*}[h]
\centering
\caption{Comparison between distributionally robust chance-constrained model with chance-constrained model with Gaussian distribution}
\begin{tabular}{lcccc}
\hline
 Confidence levels  & \multicolumn{2}{c}{95\%}      & \multicolumn{2}{c}{99\%} \\ \hline
 Models   & DRCC        & CC   & DRCC       & CC     \\ \hline
EB(MW)      & 4.0   & 4.0  & 4.0        & 4.0         \\
P2HH(MW)  & 3.5   & 3.5  & 3.5        & 3.5   \\
Annualized cost(\$)   & -1.7$\times10^5$ & -3.1$\times10^5$ & -1.1 $\times10^5$      & -2.7$\times10^5$        \\
Violation probability & 0.026   & 0.999    & 0          & 0.94      \\ \hline
\end{tabular}
\label{tab:Gaussian}
\end{table*}

To further illustrate the effectiveness of our developed distributionally robust chance-constrained model, we compare it with a chance-constrained model assuming Gaussian distribution, which models the wind forecast errors with Gaussian distribution. As a matter of fact, our forecast data show some ellipsoid-like Gaussian distribution features as can be visualized in Figure \ref{fig:DRCC}. The results are listed in Table \ref{tab:Gaussian}. Similarly, the model robustness does not chance power to heat facility sizing. More transmission grid regulating is involved in handling wind uncertainties in a more stringent model. Under both confidence levels, the profits from chance-constrained models roughly double that of distributionally robust chance-constrained models. However, the chance-constrained models perform poorly in the out-of-sample violation tests, which implies that the chance-constrained models inappropriately and insuffiently represent the wind forecast dataset. From this comparison, we conclude the effectiveness and robustness of our developed distributionally robust chance-constrained model.

\subsection{Profit Distribution}\label{rev}

In section \ref{P2H}, we verify the positive economical performance of introducing power to heat facilities to the integrated energy system in terms of system cost. In this section, we discuss the profit distribution among various stakeholders in the integrated energy system, i.e., wind power plant, CHP plant, P2HH/EB investors. Through this analysis, we found that although the system cost is reduced, the profit is distributed unevenly across the stakeholders. Thus, proper policies need to be designed to incentivize investments in power to heat facilities. Detailed analysis is expanded in the following.

\begin{figure}[t]
    \centering
    \includegraphics[width = \linewidth]{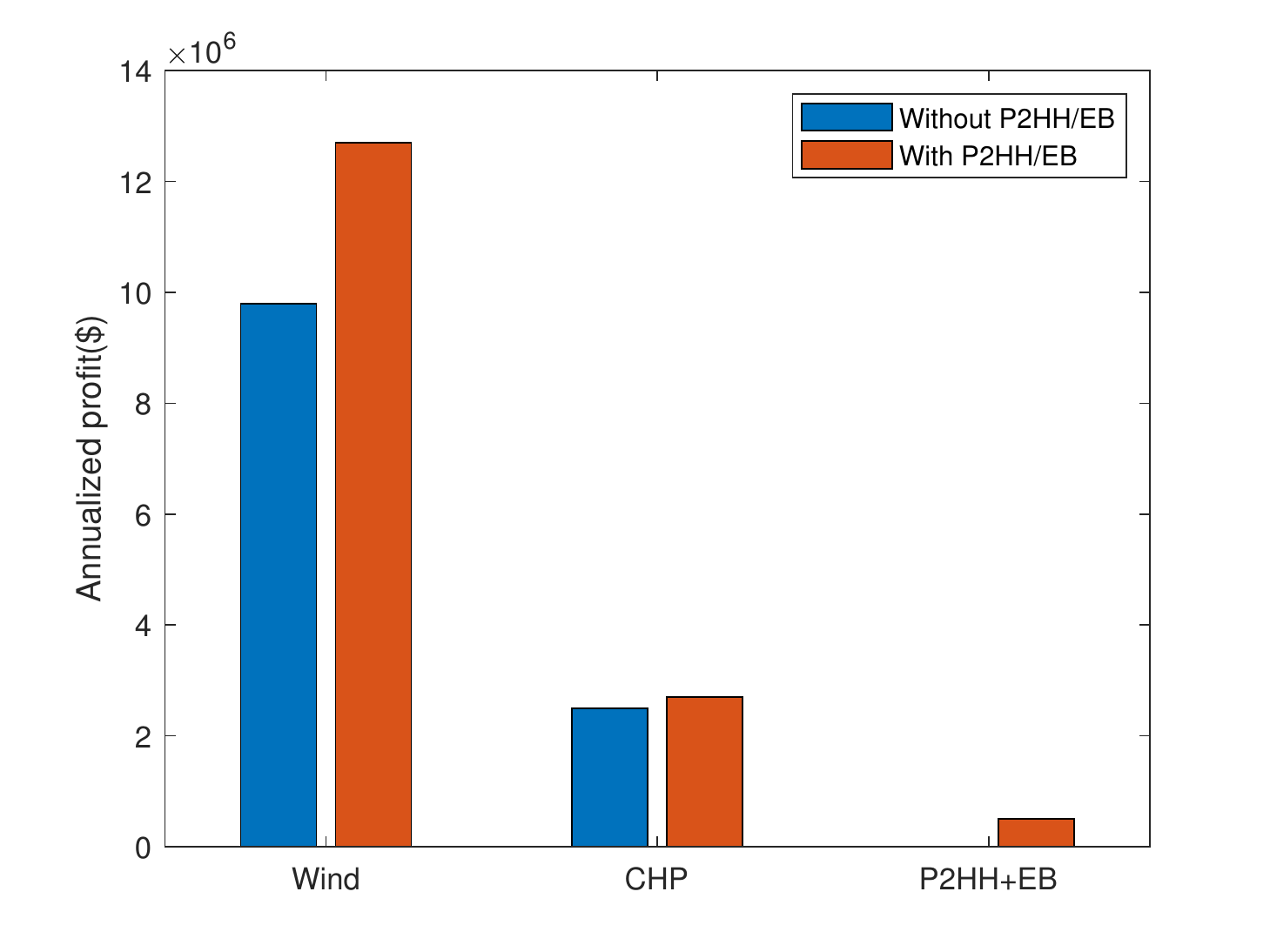}
    \caption{Profit distribution across various stakeholders}
    \label{fig:rev}
\end{figure}

We assume an electricity price of \$55/MWh and a heating price of \$45/MWh. We further assume that the transmission cost is deducted from the electricity revenue pool and the rest is split between the wind plants and the CHP plant according to their individual electricity production. Before introducing power to heat facilities, the wind plant and the CHP plants split electricity revenue while the CHP plant also takes revenue from its heat production. After introducing P2HH and electric boiler to the system, the P2HH/EB investors earn revenue from heat and hydrogen selling, while investing these facilities and purchasing electricity for self-use. The profits for each party are shown in Figure \ref{fig:rev}. The wind power plants' revenue increases as a result of high wind power utilization and lower transmission cost. Unexpectedly, the CHP plant's profit also increases even though its power and heat production decreases, which is compensated from the same reason as for the wind plant: higher power utilization and lower transmission cost. The flexibility service investor also makes profits from its investment, but at a much lower level than the wind power plants, which implies that the majority of the reduced system cost falls on the wind power plants. Therefore, additional incentive mechanism needs to be invented to incentivize investments of these flexibility resources from external investors. One possible option is to negotiate a bilateral contract between the wind power plants and the flexibility resource investors to transfer part of the incremental profit from the wind power plants to the flexibility resource investors.

On the other side, if the investment decision falls on existing participants, i.e., the wind power plants and the CHP plant, the wind power plant is more incentivized to make such investments (forming a power, heat, hydrogen energy hub) as it can increase its overall profit by 35\%. The CHP plant is more reluctant to make such investment strategies. Even though it can increase its profit by 28\%, the majority of benefits from flexible system operations fall on the wind power plants. Similarly, a proper bilateral contract between the wind power plants and the CHP plant can then be initiated to incentivize such investment from the CHP plant. 

To summarize, the wind power plants are most incentivized to make investments in power to heat facilities. The CHP plant and external investors rely on further incentives to make such investments, e.g., bilateral contracts with the wind power plants to claim part of the wind plants' extra profit.

%% file: 6.Conclusions.tex
\section{Conclusions}\label{sec:conclusions}
In this study, we investigate the use of power to heat facilities including electrolysers and electric boilers to provide additional flexibility to an integrated energy system including wind plants, combined heat and power plant as well as electric and heating loads. A distributionally robust chance-constrained model is developed to consider wind generation uncertainties, while linear decision rules are applied to simulate recourse actions. The developed model is applied in a case study to optimize electrolyser and electric boiler sizing and system operations.

By introducing the electrolyser and electric boiler into the integrated energy system, decreased system cost, improved combined heat and power plant flexibility and reduced inverse power flow are identified. Comparing the distributionally robust optimization models under different confidence levels, it is found that models with higher confidence levels though have higher costs, exhibit stronger robustness from the out-of-sample violation tests. Further comparing the developed distributionally robust optimization model with a benchmark chance-constrained model which assumes wind power forecast errors follow Gaussian distribution, we found that the distributionally robust optimization model shows better effectiveness and robustness than the chance-constrained model. Detailed profit analysis reveals that although the system cost is enhanced for the integrated energy system, the profit is distributed unevenly across various stakeholders, where the wind power plants take most of the benefits from extra system flexibility, hence is most incentivized to make such investments. The CHP plant and external investors rely on bilateral contracts with the wind power plants to be motivated to invest in flexibility resources. The findings from this study can motivate regulators to make proper policies to incentivize investments in flexibility resources and establish a more reliable power grid.

Future work can be conducted in the following areas. From the model perspective, distribution network and district heating network topology could be incorporated into the model. Electrical energy storage and heat storage could be considered. From the methodology perspective, alternatives to construct the ambiguity set such as Wasserstein distance are worth examining. Generalized decision rules could be adopted to simulate recourse actions.

%% file: 7.appendix.tex
\appendix
\section{Linear Decision Rules} \label{app:LDR}
\begin{subequations} \label{eq:affinePolicy}
\begin{equation} \label{eq:apx}
    \mathbf{\widetilde{x}}_{r,t}(\boldsymbol{\omega}_{r,t}) = \mathbf{x}_{r,t}  + (\mathbf{1}^{\top} \boldsymbol{\omega}_{r,t})  \boldsymbol{\alpha}_{r,t}, \forall r, \forall t
\end{equation}
\begin{equation}\label{eq:aptrans}
\widetilde{p}_{r,t}^{trans}(\boldsymbol{\omega}_{r,t}) = p_{r,t}^{trans}+ \beta_{r,t} (\mathbf{1}^{\top} \boldsymbol{\omega}_{r,t}) , \forall r, \forall t
\end{equation}
\begin{equation}\label{eq:apEBp}
    \widetilde{p}_{r,t}^{EB}(\boldsymbol{\omega}_{r,t}) = p_{r,t}^{EB}+ \rho_{r,t} (\mathbf{1}^{\top} \boldsymbol{\omega}_{r,t}) , \forall r, \forall t
\end{equation}
\begin{equation}\label{eq:apEBq}
    \widetilde{q}_{r,t}^{EB}(\boldsymbol{\omega}_{r,t}) = q_{r,t}^{EB}+ \nu_{r,t} (\mathbf{1}^{\top} \boldsymbol{\omega}_{r,t}) , \forall r, \forall t
\end{equation}
\begin{equation}\label{eq:aphydrogen}
    \widetilde{n}_{r,t}^{H2}(\boldsymbol{\omega}_{r,t}) = n_{r,t}^{H2} + \gamma_{r,t} (\mathbf{1}^{\top} \boldsymbol{\omega}_{r,t}), \forall r, \forall t
\end{equation}
\begin{equation}\label{eq:apP2HH}
    \widetilde{p}_{r,t}^{P2HH}(\boldsymbol{\omega}_{r,t}) = p_{r,t}^{P2HH} + \Delta_{r,t} (\mathbf{1}^{\top} \boldsymbol{\omega}_{r,t}), \forall r, \forall t
\end{equation}
\begin{equation}\label{eq:app2hh}
    \widetilde{p}_{r,t}^{p2hh}(\boldsymbol{\omega}_{r,t}) = p_{r,t}^{p2hh} + \delta_{r,t} (\mathbf{1}^{\top} \boldsymbol{\omega}_{r,t}), \forall r, \forall t
\end{equation}
\begin{equation}\label{eq:apEXC}
\widetilde{q}^{EXC}_{r,t}(\boldsymbol{\omega}_{r,t}) = q^{EXC}_{r,t} + \Lambda_{r,t} (\mathbf{1}^{\top} \boldsymbol{\omega}_{r,t}), \forall r, \forall t
\end{equation}
\begin{equation}\label{eq:apexc}
\widetilde{q}^{exc}_{r,t}(\boldsymbol{\omega}_{r,t}) = q^{exc}_{r,t} + \lambda_{r,t} (\mathbf{1}^{\top} \boldsymbol{\omega}_{r,t}), \forall r, \forall t
\end{equation}
\begin{equation}\label{eq:aphP2HH}
    \widetilde{h}_{r,t}^{P2HH}(\boldsymbol{\omega}_{r,t}) = h_{r,t}^{P2HH} + \Pi_{r,t} (\mathbf{1}^{\top} \boldsymbol{\omega}_{r,t}), \forall r, \forall t
\end{equation}
\begin{equation}\label{eq:aphp2hh}
    \widetilde{h}_{r,t}^{p2hh}(\boldsymbol{\omega}_{r,t}) = h_{r,t}^{p2hh} + \pi_{r,t} (\mathbf{1}^{\top} \boldsymbol{\omega}_{r,t}), \forall r, \forall t
\end{equation}
\begin{equation}\label{eq:apqp2hh}
    \widetilde{q}_{r,t}^{p2hh}(\boldsymbol{\omega}_{r,t}) = q_{r,t}^{p2hh} + \kappa_{r,t} (\mathbf{1}^{\top} \boldsymbol{\omega}_{r,t}), \forall r, \forall t
\end{equation}
\begin{equation} \label{eq:apy}
    \mathbf{\widetilde{y}}_{r,t}(\boldsymbol{\omega}_{r,t}) = \mathbf{y}_{r,t}  + (\mathbf{1}^{\top} \boldsymbol{\omega}_{r,t})  \boldsymbol{\zeta}_{r,t}, \forall r, \forall t
\end{equation}
\begin{equation}\label{eq:apTemp}
        \widetilde{T}_{r,t}(\boldsymbol{\omega}_{r,t}) = T_{r,t} + \mu_{r,t} (\mathbf{1}^{\top} \boldsymbol{\omega}_{r,t}) , \forall r, t\geq 2
\end{equation}
\begin{equation}\label{eq:m_tank}
        \widetilde{m}^{H2}_{r,t}(\boldsymbol{\omega}_{r,t}) = m^{H2}_{r,t} + \upsilon_{r,t} (\mathbf{1}^{\top} \boldsymbol{\omega}_{r,t}) , \forall r, \forall t
\end{equation}
\end{subequations}

\section{Model Reformulation}\label{app:reformulation}

\subsection{Objective Function}
The expectations in the objective function (\ref{eq:obj}) are derived as (\ref{eq:expection}).
\begin{subequations} \label{eq:expection}
\begin{equation}\label{eq:expCHPcost}
\begin{split}
    \mathbb{E}^{D}\bigg[\mathbf{C}^{\top} \widetilde{\mathbf{x}}_{r, t}\left(\boldsymbol{\omega}_{r, t}\right)\bigg] = \mathbb{E}^{D}\bigg[\mathbf{C}^{\top} \bigg(\mathbf{x}_{r,t}  + (\mathbf{1}^{\top} \boldsymbol{\omega}_{r,t})  \boldsymbol{\alpha}_{r,t}\bigg)\bigg]\\
    =\mathbf{C}^{\top} \mathbf{x}_{r,t} + \mathbf{C}^{\top}(\mathbf{1}^{\top} \mathbb{E}^{D}(\boldsymbol{\omega}_{r,t}))  \boldsymbol{\alpha}_{r,t} = \mathbf{C}^{\top} \mathbf{x}_{r,t}, \forall r, \forall t
\end{split}
\end{equation}   
\begin{equation}\label{eq:exptranscost}
\begin{split}
    \mathbb{E}^{D}\bigg[\left(\widetilde{p}_{r, t}^{\operatorname{trans}}\left(\boldsymbol{\omega}_{r, t}\right)\right)^{2}\bigg] = \mathbb{E}^{D}\bigg[\bigg(p_{r, t}^{t r a n s}+\beta_{r, t}\left(\mathbf{1}^{\top} \boldsymbol{\omega}_{r, t}\right)\bigg)^2\bigg]\\
    = \mathbb{E}^{D}\bigg[ \left(p_{r, t}^{t r a n s}\right)^{2}+ \beta_{r, t}^{2} \left(\mathbf{1}^{\top} \boldsymbol{\omega}_{r, t}\right)^2 + 2p_{r, t}^{t r a n s}\beta_{r, t}\left(\mathbf{1}^{\top} \boldsymbol{\omega}_{r, t}\right)\bigg] \\
    =\left(p_{r, t}^{t r a n s}\right)^{2} + \beta_{r, t}^{2} \mathbb{E}^{D}\bigg[\mathbf{1}^{\top} \boldsymbol{\omega}_{r, t} \boldsymbol{\omega}_{r, t}^{\top} \mathbf{1} \bigg] \\= \left(p_{r, t}^{t r a n s}\right)^{2} + \beta
  _{r,t}^2 \mathbf{1}^{\top}\boldsymbol{\Sigma}_{r,t}\mathbf{1}
\end{split}
\end{equation}
\end{subequations}

Based on the above derivation (\ref{eq:expection}), the \textit{max.} term and uncertainty terms in the objective function (\ref{eq:obj}) disappear as the distributions in the ambiguity set share the same first- and second-order moments (i.e., mean and covariance). The two \textit{min.} terms merge and the objective function (\ref{eq:obj}) is hence reformulated as (\ref{eq:objdeterministic}), which is a single-level quadratic program. 
\begin{equation}\label{eq:objdeterministic}
\begin{split}
  \underset{\Theta}{\min.}  \Bigg\{ c^{el} n^{el} + c^{conv}P^{conv} + c^{comp}m^{comp} +c^{tank} m^{tank} \\+ c^{EB}P^{EB}  + \sum_r k_r \sum_t \bigg[c^{su} u_{r,t}^{su} + c^{sd} u_{r,t}^{sd} \bigg ] +  \sum_r k_r \sum_t \\ \bigg[\mathbf{C}^{\top} \mathbf{x}_{r,t}  + c^{trans}\bigg[(p_{r,t}^{trans})^2 + \beta
  _{r,t}^2 \mathbf{1}^{\top}\boldsymbol{\Sigma}_{r,t}\mathbf{1}\bigg] -c^{H2} n_{r,t}^{H2}\bigg]\bigg \}  
\end{split}
\end{equation}

\subsection{Reformulating Chance Constraints}
Applying linear decision rules, the distributionally robust chance constraints can be reformulated as second-order cone constraints. Without loss of generality, we herein derive the reformulation of the following distributionally robust chance constraint (\ref{eq:DRCCgeneral}) using Cantelli's inequality (\ref{eq:cantelli}), where $\mathcal{P}$ refers to the ambiguity set grouping distributions $D$ having the same first- and second-order moment (i.e., $\boldsymbol{\mu}, \boldsymbol{\Sigma}$). $A$ refers to a vector with a proper dimension and $b$ is a scalar.
\begin{equation}\label{eq:DRCCgeneral}
     \underset{D \in \mathcal{P}}{\min} \mathbb{P}[ A^{\top} \boldsymbol{\gamma} \leq b] \geq 1 - \epsilon
\end{equation}
\begin{equation} \label{eq:cantelli}
    \mathbb{P}[X^* - E(X^*) \leq k] \geq 1-\frac{\sigma^2(X^*)}{\sigma^2(X^*)+ k^2}  (k > 0)
\end{equation}

The Cantelli's inequality (i.e., one-sided Chebyshev inequality) is stated as (\ref{eq:cantelli}), where $X^*$ refers to a random variable and k is a positive scalar. It has been shown in \citep{tightCantelli} that this bound is tight which implies that there exist a distribution with mean $E(X^*)$ and covariance $\sigma^2(X^*)$ satisfying the equality, i.e., $\mathbb{P}[X^* - E(X^*) \leq k] = 1- \frac{\sigma^2(X^*)}{\sigma^2(X^*)+ k^2}  (k > 0)$. In the following (\ref{eq:socpDerive}), we establish the equivalence between the distributionally robust chance constraint (\ref{eq:DRCCgeneral}) and a second-order cone constraint, i.e., $\left\|\mathbf{\Sigma}^{1 / 2} A\right\|_{2} \leq \sqrt{\frac{\epsilon}{1-\epsilon}}\left(b-A^{\top} \boldsymbol{\mu}\right)$ using Cantelli's inequality.

\begin{subequations}\label{eq:socpDerive}
\begin{equation}\label{eq:socpDerive1}
\begin{split}
    \min _{D \in \mathcal{P}} \mathbb{P}\left[A^{\top} \boldsymbol{\gamma} \leq b\right]  = \min _{D \in \mathcal{P}} \mathbb{P}\left[A^{\top} \boldsymbol{\gamma} - A^{\top} \boldsymbol{\mu} \leq b-A^{\top} \boldsymbol{\mu}\right] \\= 1-\frac{\sigma^2(A^{\top} \boldsymbol{\gamma})}{\sigma^2(A^{\top} \boldsymbol{\gamma})+ k^2} = 1-\frac{A^{\top} \boldsymbol{\Sigma} A}{A^{\top} \boldsymbol{\Sigma} A+ k^2} 
\end{split}
\end{equation}
\begin{equation}\label{eq:socpDerive2}
\begin{split}
    1-\frac{A^{\top} \boldsymbol{\Sigma} A}{A^{\top} \boldsymbol{\Sigma} A+ k^2} \geq 1-\epsilon \Longrightarrow \sqrt{A^{\top} \boldsymbol{\Sigma} A} \leq \sqrt{\frac{\epsilon}{1-\epsilon}} (b\\-A^{\top} \boldsymbol{\mu})  \Longrightarrow \left\|\boldsymbol{\Sigma}^{1/2} A\right\|_2 \leq \sqrt{\frac{\epsilon}{1-\epsilon}} (b-A^{\top} \boldsymbol{\mu})
\end{split}
\end{equation}
\end{subequations}

Using the established equivalence between the distributionally robust chance constraint (\ref{eq:DRCCgeneral}) and (\ref{eq:socpDerive2}), we can reformulate the afore-mentioned distributionally robust chance constraints, which are shown as below. The equality constraints which contains affine recourse actions can also be reformulated by grouping nominal terms and stochastic terms. As an example, (\ref{eq:CHPintegrity}) can be expressed as $\mathbf{1}^{\top}\mathbf{x_{r,t}} +\mathbf{1}^{\top} \boldsymbol{\alpha}_{r,t} (\mathbf{1}^{\top}\boldsymbol{\omega}_{r,t}) = u_{r,t} $ by substituting recourse actions with corresponding affine policies. By separating nominal and stochastic terms, this equality can be expressed as (\ref{eq:chanceToDxintegrity}).
\begin{subequations}
\begin{equation}\label{eq:chanceToDxintegrity}
    \mathbf{1}^{\top}\mathbf{x}_{r,t} = u_{r,t},\mathbf{1}^{\top}\boldsymbol{\alpha}_{r,t} = 0, \forall r, \forall t
\end{equation}
\begin{equation}\label{eq:chanceToDxBigger}
   \left\|\boldsymbol{\alpha}_{r,t, k}  \boldsymbol{\Sigma}_{r,t}^{1 / 2} \mathbf{1} \right\|_{2} \leq \sqrt{\frac{\epsilon}{1-\epsilon}} \mathbf{x}_{r,t,k}, \forall r, \forall t, \forall k
\end{equation}
\begin{equation}\label{eq:chanceToDxLess}
   \left\|\boldsymbol{\alpha}_{r,t, k}  \boldsymbol{\Sigma}_{r,t}^{1 / 2} \mathbf{1} \right\|_{2} \leq \sqrt{\frac{\epsilon}{1-\epsilon}} (1-\mathbf{x}_{r,t,k}), \forall r, \forall t, \forall k
\end{equation}
\begin{equation}\label{eq:chanceToDrampUp0}
   \left\| \mathbf{P}^{\top} \boldsymbol{\alpha}_{r,t} \boldsymbol{\Sigma}_{r,t}^{1 / 2} \mathbf{1}  \right\|_{2} \leq \sqrt{\frac{\epsilon}{1-\epsilon}} (SU - \mathbf{P}^{\top} \mathbf{x}_{r,t}), \forall r, t = 1
\end{equation}
\begin{equation}\label{eq:chanceToDpbalance}
\begin{split}
    p_{r,t}^{trans} + \mathbf{P}^{\top}\mathbf{x}_{r,t} + \mathbf{1}^{\top} \mathbf{m}_{r,t} =  p_{r,t}^{P2HH} / \eta^{conv} + n^{H2}_{r,t}\eta^{comp}\\+ p_{r,t}^{EB} + \mathbf{1}^{\top} \mathbf{d}^{p}_{r,t},\\ \beta_{r,t} + \mathbf{P}^{\top} \boldsymbol{\alpha}_{r,t} + 1 = \Delta_{r,t}/\eta^{conv} + \gamma_{r,t}\eta^{comp} + \rho_{r,t},  \forall r, \forall t
\end{split}
\end{equation}
\begin{equation}\label{eq:chanceToDqbalance}
\begin{split}
     \mathbf{Q}^{\top}\mathbf{x}_{r,t}  + q^{EXC}_{r,t}+ q^{EB}_{r,t} =  \mathbf{1}^{\top} \mathbf{d}^{q}_{r,t}, \\   \mathbf{Q}^{\top} \boldsymbol{\alpha}_{r,t} + \Lambda_{r,t}+\nu_{r,t} = 0, \forall r, \forall t
\end{split}
\end{equation}
\begin{equation}\label{eq:CTDEBbig}
    \left\|\rho_{r,t}  \boldsymbol{\Sigma}_{r,t}^{1 / 2} \mathbf{1} \right\|_{2} \leq \sqrt{\frac{\epsilon}{1-\epsilon}} p^{EB}_{r,t}, \forall r, \forall t
\end{equation}
\begin{equation}\label{eq:CTDEBless}
    \left\|\rho_{r,t}  \boldsymbol{\Sigma}_{r,t}^{1 / 2} \mathbf{1} \right\|_{2} \leq \sqrt{\frac{\epsilon}{1-\epsilon}} \bigg(P^{EB}-p^{EB}_{r,t}\bigg), \forall r, \forall t
\end{equation}
\begin{equation}\label{eq:CTDEBratio}
    q^{EB}_{r,t} = p^{EB}_{r,t} \eta^{EB}, \nu_{r,t} = \rho_{r,t} \eta^{EB}, \forall r, \forall t
\end{equation}

\begin{equation}\label{eq:singleTOstackpower}
    p_{r,t}^{P2HH} = n^{el} p_{r,t}^{p2hh}, \quad \Delta_{r,t} = n^{el} \delta_{r,t}, \forall r, \forall t
\end{equation}
\begin{equation}\label{eq:singleTOstackhydrogen}
    h_{r,t}^{P2HH} = n^{el} h_{r,t}^{p2hh}, \quad \Pi_{r,t} = n^{el} \pi_{r,t}, \forall r, \forall t
\end{equation}
\begin{equation}\label{eq:singleTOstackexc}
    q_{r,t}^{EXC} = n^{el} q_{r,t}^{exc}, \quad \Lambda_{r,t} = n^{el} \lambda_{r,t}, \forall r, \forall t
\end{equation}
\begin{equation}\label{eq:chanceToDexcBigger}
   \left\|\lambda_{r,t}  \boldsymbol{\Sigma}_{r,t}^{1 / 2} \mathbf{1} \right\|_{2} \leq \sqrt{\frac{\epsilon}{1-\epsilon}} q^{exc}_{r,t}, \forall r, \forall t
\end{equation}
\begin{equation}\label{eq:CTDelecPower}
   p^{p2hh}_{r,t} = h^{p2hh}_{r,t}+ q^{p2hh}_{r,t}, \delta_{r,t} = \pi_{r,t} + \kappa_{r,t}, \forall r, \forall t
\end{equation}
\begin{equation}\label{eq:hydrogenStack}
    n^{H2}_{r,t} = \frac{3.6 \times 10^6}{U_{tn}F} h_{r,t}^{P2HH}, \quad \gamma_{r,t} = \frac{3.6 \times 10^6}{U_{tn}F} \Pi_{r,t}, \forall r, \forall t
\end{equation}
\begin{equation}\label{eq:chanceToDyintegrity}
    \mathbf{1}^{\top}\mathbf{y}_{r,t} = 1,\mathbf{1}^{\top}\boldsymbol{\zeta}_{r,t} = 0, \forall r, \forall t
\end{equation}
\begin{equation}\label{eq:chanceToDyBigger}
   \left\|\boldsymbol{\zeta}_{r,t, i}  \boldsymbol{\Sigma}_{r,t}^{1 / 2} \mathbf{1} \right\|_{2} \leq \sqrt{\frac{\epsilon}{1-\epsilon}} \mathbf{y}_{r,t,i}, \forall r, \forall t, \forall i
\end{equation}
\begin{equation}\label{eq:chanceToDyLess}
   \left\|\boldsymbol{\zeta}_{r,t, i}  \boldsymbol{\Sigma}_{r,t}^{1 / 2} \mathbf{1} \right\|_{2} \leq \sqrt{\frac{\epsilon}{1-\epsilon}} (1-\mathbf{y}_{r,t,i}), \forall r, \forall t, \forall i
\end{equation}
\begin{equation}\label{eq:chanceToDhydrogenPower}
    \mathbf{P_{H2}}^{\top} \mathbf{y}_{r,t}= h^{p2hh}_{r,t}, \mathbf{P_{H2}}^{\top} \boldsymbol{\zeta}_{r,t} = \pi_{r,t},\forall r, \forall t
\end{equation}
\begin{equation}\label{eq:chanceToDheatingPower}
    \mathbf{P_{Heat}}^{\top} \mathbf{y}_{r,t} = q^{p2hh}_{r,t}, 
     \mathbf{P_{Heat}}^{\top} \boldsymbol{\zeta}_{r,t} = \kappa_{r,t}, \forall r, \forall t
\end{equation}
\begin{equation}\label{eq:ChanceToDtemperature0}
    \mathbf{T}^{\top} \mathbf{y}_{r,t} = T_{r,t}, \mathbf{T}^{\top} \boldsymbol{\zeta}_{r,t} = 0, \forall r, t = 1
\end{equation}
\begin{equation}\label{eq:ChanceToDtemperature}
    \mathbf{T}^{\top} \mathbf{y}_{r,t} = T_{r,t}, \mathbf{T}^{\top} \boldsymbol{\zeta}_{r,t} = \mu_{r,t},\forall r,  t \geq 2
\end{equation}
\begin{equation}\label{eq:chanceToDTempBigger}
   \left\|\mu_{r,t}  \boldsymbol{\Sigma}_{r,t}^{1 / 2} \mathbf{1} \right\|_{2} \leq \sqrt{\frac{\epsilon}{1-\epsilon}} \bigg(T_{r,t} - T_{min}\bigg), \forall r, t \geq 2
\end{equation}
\begin{equation}\label{eq:chanceToDTempLess}
   \left\|\mu_{r,t}  \boldsymbol{\Sigma}_{r,t}^{1 / 2} \mathbf{1} \right\|_{2} \leq \sqrt{\frac{\epsilon}{1-\epsilon}} \bigg(T_{max} - T_{r,t} \bigg), \forall r, t \geq 2
\end{equation}
\begin{equation}\label{eq:chanceToDTemp24Bigger}
\begin{split}
   \left\|\Bigg( (1-\frac{1}{R^{eqv}C})\mu_{r,t} + \frac{1}{C}\kappa_{r,t} - \frac{1}{C}\lambda_{r,t} \Bigg)  \boldsymbol{\Sigma}_{r,t}^{1 / 2} \mathbf{1} \right\|_{2}\\ \leq \sqrt{\frac{\epsilon}{1-\epsilon}}  \bigg((1-\frac{1}{R^{eqv}C})T_{r,t}  \\+ \frac{1}{C}(q^{p2hh}_{r,t} - q^{exc}_{r,t} + \frac{1}{R^{eqv}}T_a)- T_{min}\bigg), \forall r, t =|t|
\end{split}
\end{equation}
\begin{equation}\label{eq:chanceToDTemp24Less}
\begin{split}
   \left\|\Bigg( (1-\frac{1}{R^{eqv}C})\mu_{r,t} + \frac{1}{C}\kappa_{r,t} - \frac{1}{C}\lambda_{r,t} \Bigg)  \boldsymbol{\Sigma}_{r,t}^{1 / 2} \mathbf{1} \right\|_{2} \\\leq \sqrt{\frac{\epsilon}{1-\epsilon}}  \bigg(-(1-\frac{1}{R^{eqv}C})T_{r,t} \\  - \frac{1}{C}(q^{p2hh}_{r,t} - q^{exc}_{r,t} + \frac{1}{R^{eqv}}T_a) + T_{max}\bigg), \forall r, t =|t|
\end{split}
\end{equation}
\end{subequations}
\begin{subequations}\label{eq:new}
\begin{equation}\label{eq:CTDm_tank1}
    m^{H2}_{r,t} = n^{H2}_{r,t}, \upsilon_{r,t} = \gamma_{r,t}, \forall r, t = 1
\end{equation}
\begin{equation}\label{eq:CTDh2end}
   \left\|\upsilon_{r,t}  \boldsymbol{\Sigma}_{r,t}^{1 / 2} \mathbf{1} \right\|_{2} \leq \sqrt{\frac{\epsilon}{1-\epsilon}} \bigg(m^{tank} - m^{H2}_{r,t}\bigg), \forall r, t = |t|   
\end{equation}
\begin{equation}\label{eq:CTDconverter}
   \left\|\Delta_{r,t}  \boldsymbol{\Sigma}_{r,t}^{1 / 2} \mathbf{1} \right\|_{2} \leq \sqrt{\frac{\epsilon}{1-\epsilon}} \bigg(P^{conv}\eta^{conv} - p^{P2HH}_{r,t}\bigg), \forall r, \forall t   
\end{equation}
\begin{equation}\label{eq:CTDcomp}
   \left\|\gamma_{r,t}  \boldsymbol{\Sigma}_{r,t}^{1 / 2} \mathbf{1} \right\|_{2} \leq \sqrt{\frac{\epsilon}{1-\epsilon}} \bigg(m^{comp} - n^{H2}_{r,t}\bigg), \forall r, \forall t 
\end{equation}
\end{subequations}

\subsection{Inter-temporal Constraints}
Constraints (\ref{eq:rampup2}) and (\ref{eq:rampdown2}) are a set of inter-temporal constraints, which involves uncertainties in two consecutive hours. Similar to \citep{Pourahmadi2019}, the uncertainty parameter vector is set to include the two consecutive hours, i.e., $\left[\begin{array}{c}\omega_{r, t-1} \\ \omega_{r, t}\end{array}\right] \in \mathbb{R}^{2Z}$ for these constraints. (\ref{eq:rampup2}) and (\ref{eq:rampdown2}) are hence reformulated as (\ref{eq:rampupReformulation}) and (\ref{eq:rampdownReformulation}) respectively. In order to establish the corresponding second-order cone constraints, a new covariance matrix $\boldsymbol{\Sigma}_{r(t-1), r t}\in \mathbb{R}^{2 Z \times 2 Z}$ is defined in (\ref{eq:newCovariance}) to model not only the spatial correlation but also the inter-temporal correlation of forecast errors. Using this new covariance matrix, (\ref{eq:rampupReformulation}) and (\ref{eq:rampdownReformulation}) can be equivalently reformulated as second order cone constraints (\ref{eq:chanceToDrampUp}) and (\ref{eq:chanceToDrampDown}).

\begin{equation}\label{eq:newCovariance}
    \boldsymbol{\Sigma}_{r(t-1),rt} = \left[\begin{array}{cc}
\boldsymbol{\Sigma}_{r(t-1)} & \boldsymbol{\Upsilon}_{r(t-1),rt} \\
\boldsymbol{\Upsilon}_{r(t-1),rt} & \boldsymbol{\Sigma}_{rt}
\end{array}\right]
\end{equation}
\begin{subequations}
\begin{equation} \label{eq:rampupReformulation}
\begin{split}
   \underset{D \in \mathcal{P}_{r(t-1),rt}}{\min} \mathbb{P}\Bigg[ \mathbf{P}^{\top} \boldsymbol{\alpha}_{r,t} \left[\begin{array}{c}
\mathbf{0} \\
\mathbf{1}
\end{array}\right]^{\top} \left[\begin{array}{c}
\boldsymbol{\omega}_{r,t-1} \\
\boldsymbol{\omega}_{r,t}
\end{array}\right]- \mathbf{P}^{\top} \boldsymbol{\alpha}_{r,t-1} \left[\begin{array}{c}
\mathbf{1} \\
\mathbf{0}
\end{array}\right]^{\top} \left[\begin{array}{c}
\boldsymbol{\omega}_{r,t-1} \\
\boldsymbol{\omega}_{r,t}
\end{array}\right] \\ \leq \mathbf{P}^{\top} \mathbf{x}_{r,t-1} - \mathbf{P}^{\top} \mathbf{x}_{r,t} +  SU (1-u_{r,t-1}) + RU u_{r,t-1} \Bigg] \geq 1 - \epsilon, \forall r, t\geq 2
\end{split}
\end{equation}
\begin{equation} \label{eq:rampdownReformulation}
\begin{split}
   \underset{D \in \mathcal{P}_{r(t-1),rt}}{\min} \mathbb{P}\Bigg[ \mathbf{P}^{\top} \boldsymbol{\alpha}_{r,t-1} \left[\begin{array}{c}
\mathbf{1} \\
\mathbf{0}
\end{array}\right]^{\top} \left[\begin{array}{c}
\boldsymbol{\omega}_{r,t-1} \\
\boldsymbol{\omega}_{r,t}
\end{array}\right] - \mathbf{P}^{\top} \boldsymbol{\alpha}_{r,t} \left[\begin{array}{c}
\mathbf{0} \\
\mathbf{1}
\end{array}\right]^{\top} \left[\begin{array}{c}
\boldsymbol{\omega}_{r,t-1} \\
\boldsymbol{\omega}_{r,t}
\end{array}\right]  \\ \leq -\mathbf{P}^{\top} \mathbf{x}_{r,t-1} + \mathbf{P}^{\top} \mathbf{x}_{r,t} +  SD (1-u_{r,t}) + RD u_{r,t} \Bigg] \geq 1 - \epsilon, \forall r, t\geq 2
\end{split}
\end{equation}
\begin{equation}\label{eq:chanceToDrampUp}
\begin{split}
   \left\| \boldsymbol{\Sigma}_{r(t-1),rt}^{1 / 2} \Bigg(\mathbf{P}^{\top} \boldsymbol{\alpha}_{r,t} \left[\begin{array}{c}
\mathbf{0} \\
\mathbf{1}
\end{array}\right] - \mathbf{P}^{\top} \boldsymbol{\alpha}_{r,t-1} \left[\begin{array}{c}
\mathbf{1} \\
\mathbf{0}
\end{array}\right]\Bigg)  \right\|_{2}  \leq \sqrt{\frac{\epsilon}{1-\epsilon}}\\ \Bigg(\mathbf{P}^{\top} \mathbf{x}_{r,t-1} - \mathbf{P}^{\top} \mathbf{x}_{r,t} +  SU (1-u_{r,t-1}) + RU u_{r,t-1}\Bigg), \forall r, t \geq 2
\end{split}
\end{equation}
\begin{equation}\label{eq:chanceToDrampDown}
\begin{split}
   \left\| \boldsymbol{\Sigma}_{r(t-1),rt}^{1 / 2} \Bigg(-\mathbf{P}^{\top} \boldsymbol{\alpha}_{r,t} \left[\begin{array}{c}
\mathbf{0} \\
\mathbf{1}
\end{array}\right] + \mathbf{P}^{\top} \boldsymbol{\alpha}_{r,t-1} \left[\begin{array}{c}
\mathbf{1} \\
\mathbf{0}
\end{array}\right]\Bigg)  \right\|_{2}  \leq \sqrt{\frac{\epsilon}{1-\epsilon}}\\ \Bigg(-\mathbf{P}^{\top} \mathbf{x}_{r,t-1} + \mathbf{P}^{\top} \mathbf{x}_{r,t} +  SD (1-u_{r,t}) + RD u_{r,t}\Bigg), \forall r, t \geq 2
\end{split}
\end{equation}

The temperature evolution equality constraints (\ref{eq:tempEvolution0})-(\ref{eq:tempEvolution}) and hydrogen content evolution constraint (\ref{eq:tankt2}) are also reformulated by grouping nominal and stochastic terms. Due to existence of uncertainties for two consecutive hours in these constraints, an assumption is made that the aggregated forecast error in two consecutive hours are similar, i.e., $\mathbf{1}^{\top}\boldsymbol{\omega}_{t-1} \approx \mathbf{1}^{\top}\boldsymbol{\omega}_{t}$. Using this assumption, the stochastic terms for two consecutive hours can be eliminated simultaneously, shown as (\ref{eq:CTDtempEvole0})-(\ref{eq:CTDm_tank2}).

\begin{equation}\label{eq:CTDtempEvole0}
    T_{r,t+1} = T_{r,t} + \frac{1}{C} \Bigg(q^{p2hh}_{r,t} - q^{exc}_{r,t} -  \frac{1}{R^{eqv}}(T_{r,t} - T_a) \Bigg), \forall r, t = 1
\end{equation}
\begin{equation}\label{eq:CTDtempEvole1}
    \mu_{r,t+1} =\frac{1}{C} \Bigg(\kappa_{r,t} - \lambda_{r,t} \Bigg), \forall r, t = 1
\end{equation}
\begin{equation}\label{eq:CTDtempEvole2}
    T_{r,t+1} = T_{r,t} + \frac{1}{C} \Bigg(q^{p2hh}_{r,t} - q^{exc}_{r,t} -  \frac{1}{R^{eqv}}(T_{r,t} - T_a) \Bigg), \forall r, 2 \leq t \leq |t|-1
\end{equation}
\begin{equation}\label{eq:CTDtempEvole3}
    \mu_{r,t+1} =\mu_{r,t} + \frac{1}{C} \Bigg(\kappa_{r,t} - \lambda_{r,t} - \frac{1}{R^{eqv}} \mu_{r,t} \Bigg), \forall r, 2 \leq t \leq |t|-1
\end{equation}
\begin{equation}\label{eq:CTDm_tank2}
    m^{H2}_{r,t} = m^{H2}_{r,t-1} + n^{H2}_{r,t}, \upsilon_{r,t} =\upsilon_{r,t-1} + \gamma_{r,t}, \forall r, t \geq 2
\end{equation}
\end{subequations}

\subsection{Linearizing bilinear terms}
Bilinear terms (e.g., $n^{el}p^{p2hh}_{r,t}$) exist in (\ref{eq:singleTOstackpower})-(\ref{eq:singleTOstackexc}), which cannot be handled by optimization solvers directly. Using the \textit{big M} method, these bilinear terms can be linearized as (\ref{eq:bilinear}), where (\ref{eq:binary}) defines the integer variable $n^{el}$ using a set of binary variables. As an example, (\ref{eq:ex1}) expressed (\ref{eq:p2hhstackp}) using the above binary variables, where $e_{r, t, j} = z_j p_{r, t}^{p 2 h h}$ which is further linearized in (\ref{eq:ex2}) and (\ref{eq:ex3}) using the \textit{big M} method.

\begin{subequations}\label{eq:bilinear}
\begin{equation}\label{eq:binary}
    n^{el} = \sum_{j = 1}^N 2^{j-1} z_j, z_j \in \{0,1\}, n^{el} \leq 2^N - 1
\end{equation}
\begin{equation} \label{eq:ex1}
    p_{r,t}^{P2HH} = \sum_{j = 1}^N 2^{j-1} e_{r,t,j}, \forall r, \forall t
\end{equation}
\begin{equation}\label{eq:ex2}
    - M z_j  \leq  e_{r,t,j} \leq M z_j,  \forall r, \forall t, \forall j
\end{equation}
\begin{equation}\label{eq:ex3}
    - M (1-z_j) + p_{r,t}^{p2hh} \leq  e_{r,t,j} \leq M (1-z_j) + p_{r,t}^{p2hh},  \forall r, \forall t, \forall j
\end{equation}
\end{subequations}

In summary, the overall mathematical model aims to optimize the objective function (\ref{eq:objdeterministic}) under the constraints \{(\ref{eq:facility}), (\ref{eq:integrity})-(\ref{eq:min_off}), (\ref{eq:chanceToDxintegrity})-(\ref{eq:chanceToDrampUp0}), (\ref{eq:chanceToDrampUp})(\ref{eq:chanceToDrampDown}), (\ref{eq:chanceToDpbalance})-(\ref{eq:CTDEBratio}), (\ref{eq:binary})-(\ref{eq:ex3}), (\ref{eq:chanceToDexcBigger})-(\ref{eq:chanceToDTempLess}), (\ref{eq:CTDtempEvole0})-(\ref{eq:CTDtempEvole3}), (\ref{eq:chanceToDTemp24Bigger})(\ref{eq:chanceToDTemp24Less}), (\ref{eq:CTDm_tank1}), (\ref{eq:CTDm_tank2}), (\ref{eq:CTDh2end})-(\ref{eq:CTDcomp})\}, which forms a mixed-integer second-order cone constrained quadratic program (MISOCCQP). Solvers like Gurobi can be applied to solve it.

\section{Benchmark Model: Chance-Constrained Planning with Gaussian Distribution}\label{app:gaussian}
As a general case, (\ref{eq:Gaussian}) represents a chance constraint where the uncertainty parameter $\boldsymbol{\gamma}$ follows Gaussian distribution with mean $\boldsymbol{\mu}$ and covariance $\boldsymbol{\Sigma}$. The derivation of the corresponding second-order cone constraint is seen in (\ref{eq:GaussianDerive}), which establishes the equivalence between the chance constraint (\ref{eq:Gaussian}) and the second-order cone constraint $ \left\|\boldsymbol{\Sigma}^{1/2} A\right\|_2 \leq \frac{1}{\Phi^{-1}(1-\epsilon)}(b-A^{\top} \boldsymbol{\mu})$ where $\Phi$ refers to the cumulative distribution function (CDF) of the standard Gaussian distribution and $\Phi^{-1}$ refers to its inverse function.
\begin{equation}\label{eq:Gaussian}
\mathbb{P}\left[A^{\top} \boldsymbol{\gamma} \leq b\right] \geq 1-\epsilon
\end{equation}
\begin{equation}\label{eq:GaussianDerive}
\begin{split}
    \mathbb{P}\left[A^{\top} \boldsymbol{\gamma} \leq b\right] =  \Phi\left[\frac{b-A^{\top} \boldsymbol{\mu}}{\sqrt{A^{\top}\boldsymbol{\Sigma}A} } \right]  \geq 1-\epsilon \Longrightarrow  \frac{b-A^{\top} \boldsymbol{\mu}}{\sqrt{A^{\top}\boldsymbol{\Sigma}A} } \geq \Phi^{-1}(1-\epsilon) \\  \Longrightarrow \left\|\boldsymbol{\Sigma}^{1/2} A\right\|_2 \leq \frac{1}{\Phi^{-1}(1-\epsilon)}(b-A^{\top} \boldsymbol{\mu})
\end{split}
\end{equation}

It is straightforward to identify that the second-order cone reformulations of distributionally robust chance constraints and chance constraints assuming Gaussian distribution take similar forms, with the only difference lying on the right side of the inequalities, being $\sqrt{\frac{\epsilon}{1-\epsilon}}$ and $\frac{1}{\Phi^{-1}(1-\epsilon)}$ for distributionally robust chance constraints and chance constraints with Gaussian distribution respectively. 
Obviously, distributionally robust chance constraints form a much tighter reformulation, with $\sqrt{\frac{\epsilon}{1-\epsilon}}$ being 2-5 times smaller compared
to $\frac{1}{\Phi^{-1}(1-\epsilon)}$ under different $\epsilon$, which implies distributionally robust chance-constrained planning would result in a more conservative planning result.
